\documentclass[fleqn,10pt]{wlscirep}
\usepackage[utf8]{inputenc}
\usepackage[T1]{fontenc}
\usepackage{appendix}
\usepackage{caption}    
\usepackage{ragged2e}  
\usepackage{pifont}
\usepackage{fontawesome}
\usepackage[left]{lineno} 
\usepackage[colorlinks=true, allcolors=blue]{hyperref}
\title{Geological Everything Model 3D: A Promptable Foundation Model for Unified and Zero-Shot Subsurface Understanding}
\author[1]{Yimin Dou}
\author[1,*]{Xinming Wu}
\author[2]{Nathan L Bangs}
\author[3]{Harpreet Singh Sethi}
\author[1]{Jintao Li}
\author[1]{Hang Gao}
\author[1]{Zhixiang Guo}

\affil[1]{University of Science and Technology of China, School of Earth and Space Sciences, China}
\affil[2]{University of Texas at Austin, UT Institute for Geophysics, United States}
\affil[3]{NVIDIA, United States}

\affil[*]{Corresponding author: xinmwu@ustc.edu.cn \par \faGithub~\textbf{Project page}: \url{https://douyimin.github.io/GEM}}

\begin{abstract}
Understanding Earth’s subsurface is critical for energy transition, natural hazard mitigation, and planetary science. Yet subsurface analysis remains fragmented, with separate models required for structural interpretation, stratigraphic analysis, geobody segmentation, and physical property modeling—each tightly coupled to specific data distributions and task formulations. We introduce the Geological Everything Model 3D (GEM), a physics-informed, promptable foundation model that reformulates these tasks as conditional inference along latent structural frameworks derived from subsurface imaging. By embedding geological and physical priors into its generative architecture, GEM integrates domain knowledge with machine learning to ensure structurally consistent and geologically plausible outputs. This formulation moves beyond task-specific models by enabling a shared inference mechanism, where GEM propagates human-provided prompts—such as well logs, masks, or structural sketches—through inferred structural frameworks to produce coherent subsurface interpretations. Through this mechanism, GEM achieves zero-shot generalization across tasks and modalities with heterogeneous prompt types, without retraining for new tasks or data sources. This capability emerges from a two-stage training process that combines large-scale self-supervised representation learning on over 500 field seismic volumes with adversarial fine-tuning using mixed prompts and labels across structural, stratigraphic, geobody, and property domains. GEM demonstrates broad applicability across surveys and tasks, including Martian radar stratigraphy analysis, structural interpretation in subduction zones, seismic stratigraphic interpretation, geobody delineation, and physical property modeling. By bridging expert knowledge with physics-informed generative reasoning in a structurally aware manner, GEM lays the foundation for scalable, human-in-the-loop geophysical AI—transitioning from fragmented workflows to a vertically integrated, promptable reasoning system.
\end{abstract}

\begin{document}

\flushbottom
\maketitle
\thispagestyle{empty}

\section{Introduction}\label{sec1}

Subsurface understanding is foundational to some of society's most pressing challenges: sustainable energy exploration \cite{yu2024crustal,liu2025natural}, geological carbon storage \cite{zhang2024feasibility,creasy2024co2}, natural hazard prediction \cite{wang2024earthquake,li2023earthquake} and stratigraphy  \cite{li2022layered,zhang2024buried}. 
In recent years, Artificial Intelligence (AI) has emerged as a transformative paradigm for subsurface analysis \cite{cui2023similar,bergen2019machine,clubb2023himalayan,mousavi2022deep,laurenti2024probing}. However, current implementations remain fragmented, brittle, and narrowly scoped. Most workflows depend on task-specific models---each trained on narrow datasets and bound to a single survey region---to perform  structural interpretation \cite{wu2019faultseg3d,wu2023mtl,gao2021fault}, stratigraphic analysis\cite{alaudah2019machine,gao2024optimizing,yang2023multi}, geobody segmentation \cite{xu20243d,muller2024deep,yang2024salt3dnet,gao2021channelseg3d} or physical property modeling \cite{yu2021attention,saad2025f,wu2025does,li2023self}. This fragmentation hinders generalization and requires costly retraining for every new task or area. While foundation models have revolutionized vision and language\cite{brown2020language,touvron2023llama,bi2024deepseek,kirillov2023segment,feichtenhofer2022masked}, no such model has emerged for subsurface imaging. This is not simply a matter of scale, but of epistemology: geological imaging data is sparse, indirect, and structurally entangled---lacking the high-density semantic cues that foundation models in other fields rely on. Bridging this domain gap calls for rethinking how we represent, prompt, and reason about the planetary interior---not as a patchwork of disconnected labels, but as a unified structure-conditioned generative process.

Subsurface modeling poses challenges that go beyond conventional data-driven learning. First, geological structures exhibit high spatial variability across basins, stratigraphic settings, and tectonic regimes, making it difficult for models trained on one dataset to generalize to others \cite{alaudah2019machine,gao2021channelseg3d,gao2024optimizing,xu20243d}. Second, unlike natural images, subsurface imaging  volumes provide only indirect, often noisy reflections of the underlying subsurface. These data are low in semantic density and rich in structural ambiguity, shaped by complex depositional histories and imaging artifacts. As a result, similar patterns may represent different geological features—or vice versa—creating severe non-uniqueness that hampers robust pattern recognition. Figure \ref{fig06} (a, b) illustrates this mismatch between information sparsity and geological complexity. Together, these properties produce a domain where conventional AI struggles: data is abundant in size but impoverished in meaning, and geological reasoning requires structural inference beyond what traditional models can capture.

To address these challenges, current efforts follow two main directions. One emphasizes scaling up training data—either through large-scale labeling of field seismic data or synthetic data generation across geologically diverse scenarios \cite{alaudah2019machine,di2020accelerating}. However, field data suffers from low signal quality and incomplete annotations, while synthetic volumes often fail to capture the structural complexity and stochastic variability of real geology. As illustrated in Figure \ref{fig06} (c), this trade-off between quality and diversity has hindered the development of truly generalizable models. The other direction seeks to incorporate prior knowledge, including physics-based constraints \cite{schuster2024review,chen2021seismic,wu2023sensing}, stratigraphic simulation \cite{heir2024inversion,tilke2022stratigraphic,yang2023multi}, or expert-guided prompts \cite{kanfar2025intelligent,gao2024foundation}. Yet formalizing such priors remains challenging: expert reasoning is often tacit and context-specific, and most models are not designed to integrate diverse forms of human input in a unified way. As shown in Figure \ref{fig06} (d), these two routes—data-centric scaling and prior-informed modeling—each address part of the problem, but neither overcomes the core epistemic limitations of subsurface imaging.

Several recent models have made progress toward more adaptable subsurface interpretation \cite{liu2024foundation,sheng2025workflow}. The Seismic Foundation Model (SFM), based on masked autoencoders, achieves strong cross-survey results on horizon and fault detection, but remains restricted to dense supervision and narrow task formats \cite{feichtenhofer2022masked,sheng2024seismic}. The Vision Foundation Model (VFM), adapted from the Segment Anything Model (SAM), enables prompt-guided segmentation using sparse masks or points, yet it is limited to 2D slices and requires task-specific fine-tuning for most applications \cite{kirillov2023segment,gao2024foundation}. Other methods have incorporated domain priors—such as physics-informed loss functions  \cite{schuster2024review,chen2021seismic,wu2023sensing}, stratigraphic simulation constraints \cite{heir2024inversion,tilke2022stratigraphic,yang2023multi}, and weak supervision strategies \cite{dou2021attention,xu20243d}—but struggle to accommodate heterogeneous prompt types or task modalities within a shared architecture. Moreover, most models view interpretation as per-voxel classification or segmentation, lacking a structural representation of geology that can support consistent reasoning across scales and tasks. These limitations reveal the absence of a unified, promptable, and structure-aware modeling paradigm for subsurface interpretation.

In response to these limitations, we introduce the Geological Everything Model 3D (GEM), a unified generative framework that reformulates subsurface interpretation as prompt-conditioned inference over latent structural representations. Instead of treating tasks like structural interpretation, geobody segmentation, or physical property modeling as separate problems requiring bespoke models, GEM abstracts them as different forms of dimensional completion guided by human-provided prompts—such as well logs, masks, or structural sketches. At its core, GEM learns to propagate these prompts through a structurally coherent latent space derived from the input subsurface volume, enabling geologically consistent outputs across diverse task types. This formulation unifies a wide range of interpretation goals within a single model, while maintaining flexibility for expert interaction and scalability across surveys. By shifting the modeling paradigm from task-specific classification to structure-aware generation, GEM offers a foundation for zero-shot, prompt-driven, human-in-the-loop geological reasoning.

GEM is trained in two stages to enable broad generalization without task-specific retraining. First, a self-supervised pretraining stage leverages masked representation learning over more than 500 field seismic volumes to learn structure-sensitive features without requiring labels. Then, adversarial fine-tuning is performed using synthetic data with diverse prompt-label pairs across structural, stratigraphic, geobody, and physical property domains. This combination allows GEM to align prompts with latent geological frameworks, supporting flexible input types—including  lines, masks, and well logs—and producing task-consistent outputs with minimal supervision. Once trained, GEM generalizes across surveys, data modalities, and interpretation goals, achieving zero-shot performance without architectural modification or additional fine-tuning.
We evaluate GEM across a diverse set of tasks and geological settings, demonstrating broad applicability and strong zero-shot generalization. Without retraining, GEM performs high-resolution Martian radar stratigraphy analysis, interprets complex 3D fault systems in subduction zones, conducts full seismic stratigraphic interpretation, segments geobodies such as salt domes and channels, and models properties across surveys with sparse well control. These applications span planetary exploration, structural geology and physical property modeling—tasks that traditionally require disconnected models, domain-specific workflows, and extensive retraining at each stage. In contrast, GEM consolidates structural interpretation, stratigraphic analysis, geological modeling, and property prediction within a single generative framework. This unified approach eliminates the need for task-specific pipelines and mitigates error accumulation across stages, offering a more stable, scalable, and interpretable alternative. By integrating expert prompts with structure-aware reasoning, GEM bridges geoscientific insight and generative modeling, establishing a new paradigm for human-guided subsurface AI.
This work makes three key contributions:
\begin{itemize}
	\item First, it proposes a unified generative mechanism for subsurface interpretation by reframing structural delineation, full stratigraphic analysis, geobody segmentation, and physical property modeling as prompt-guided completions over latent geological frameworks derived from imaging data. This formulation enables structure-aware reasoning across tasks within a shared representation space.
	
	\item Second, it introduces GEM, a foundation model trained through self-supervised representation learning followed by adversarial fine-tuning with mixed prompts and labels. GEM supports flexible expert interaction by integrating heterogeneous human prompts—such as masks, sketches, and well logs—into coherent, geologically consistent outputs, enabling human-in-the-loop control and interpretation without task-specific retraining.
	
	\item Third, it demonstrates GEM 's strong zero-shot generalization and broad applicability across surveys, modalities, and interpretation goals, including Martian radar stratigraphy, 3D fault interpretation in subduction zones, full seismic stratigraphic interpretation, geobody segmentation, and physical property modeling with sparse well control—often superior to task-specific supervised models. 
\end{itemize}
Together, these contributions establish a new foundation for prompt-driven, human-in-the-loop, and generalizable geophysical AI.

\section{Results}
\subsection{GEM: A Unified and Prompt-Driven Reasoning Framework for Subsurface imaging}
\begin{figure}[!h]
	\includegraphics[scale=0.75]{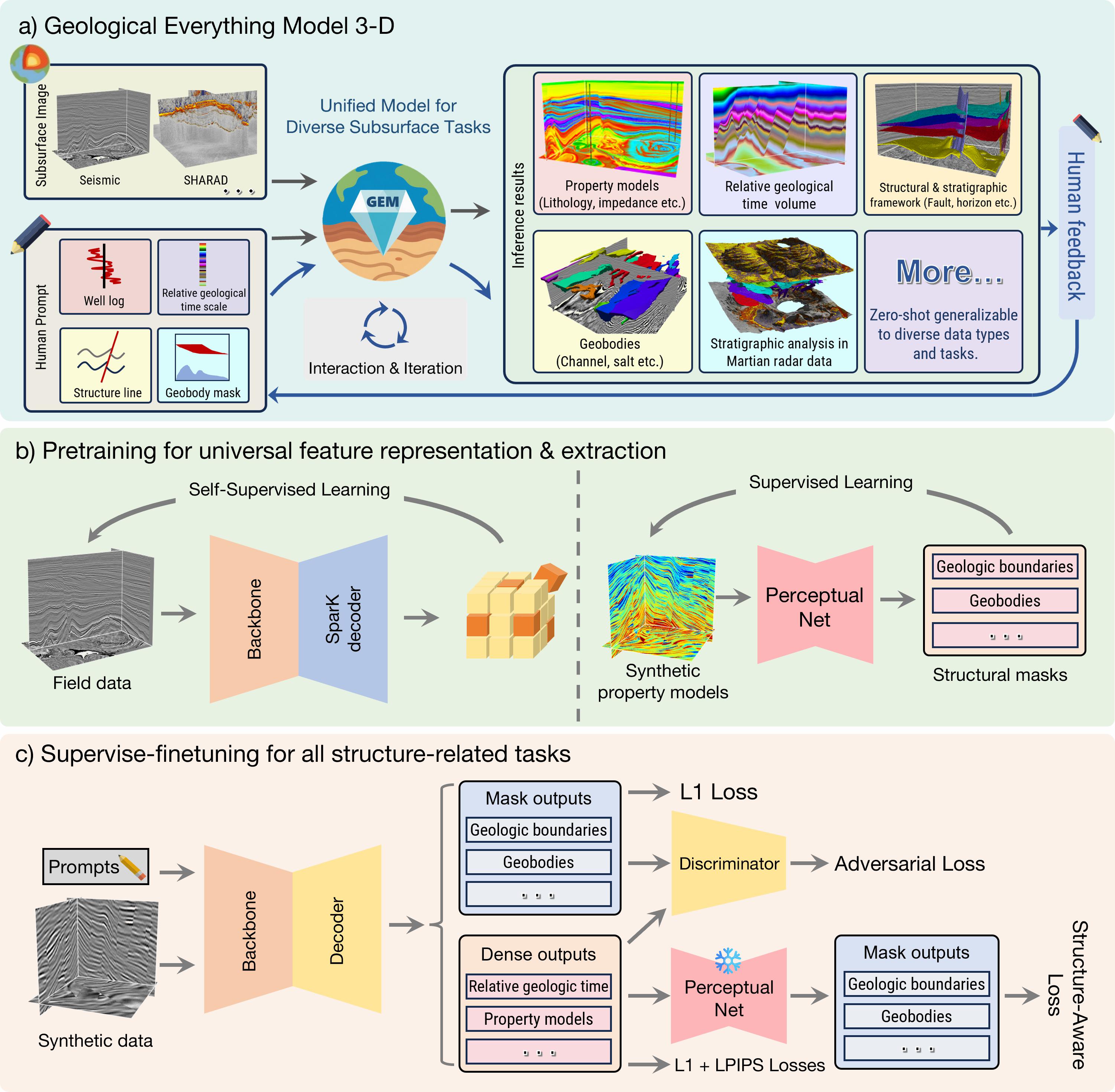}
	\centering\caption{Overview of GEM-3D, a promptable foundation model for unified and zero-shot subsurface modeling. (a) GEM supports a wide range of subsurface imaging tasks, including structural and stratigraphic interpretation (e.g., faults, horizons, unconformities, relative geological time), geobody segmentation (e.g., channels, salt bodies), and physical property modeling (e.g., impedance, gamma ray, lithology). By conditioning on diverse human prompts—such as well logs, structural or stratigraphic sketches, and masks—GEM generates geologically coherent 3D results with support for expert interaction and iterative refinement. (b) GEM is trained in two stages. During pretraining, it learns structure-sensitive representations through masked modeling on large-scale field seismic data using a sparse decoder. In parallel, a perception network is trained on synthetic structural masks and property volumes to extract high-level geological features for structure-aware supervision in the next stage. (c) In the fine-tuning stage, GEM is optimized using adversarial and perceptual losses across diverse output types. Prompts guide the generative process, while a structure-aware discriminator enforces geological plausibility. Once trained, GEM generalizes across tasks, data types, and geological regions in a zero-shot manner.}
	\label{fig01}
\end{figure}

GEM is designed as a unified, prompt-driven foundation model for diverse subsurface imaging tasks. It reformulates structural interpretation, stratigraphic analysis, geobody segmentation, and physical property modeling as conditional generative processes, driven by sparse user-provided prompts and executed over a shared latent geological framework. This formulation allows GEM to handle multiple interpretation goals within a single architecture. As illustrated in Figure 1a, GEM supports a broad spectrum of tasks—including fault and horizon extraction, relative geological time (RGT) estimation, geobody segmentation (e.g., channels, salt, volcanic units), and physical property modeling (e.g., impedance, gamma ray, lithology)—all through a consistent, prompt-conditioned inference mechanism.

A defining feature of GEM is its interactive and structure-aware reasoning paradigm. Unlike conventional models that rely on static input-output mappings, GEM supports flexible prompting over input subsurface images. Experts can provide sparse inputs—such as masks, well logs, or structural sketches—and iteratively refine outputs in real time. For example, in structural interpretation or geobody segmentation, users can mark key boundaries or features on a 2D section to guide 3D inference, then adjust unsatisfactory regions through pixel-level corrections. For RGT estimation, GEM takes a vertically aligned RGT scale prompt linearly scaled from 0 to 1, and iteratively refines the prediction by injecting new prompts at regions with high misalignment (Figure \ref{fig12}). In physical property modeling, sparse well logs serve as conditioning prompts to generate geologically consistent volumes. These multimodal prompts are propagated through the latent structural framework inferred from an input subsurface image, enabling consistent and context-aware generation across tasks. Appendices Figure \ref{fig11} and accompanying videos demonstrate this interactive workflow, where experts visualize, intervene, and re-infer outputs in a seamless loop. This human-in-the-loop capability allows GEM to integrate expert knowledge fluidly, improving interpretability, adaptability, and alignment with real-world geoscience practice.

To enable this unified and interactive inference, GEM is trained by a two-stage strategy (Figure \ref{fig01} (b, c)). During self-supervised pretraining, the model learns structure-sensitive representations from over 500 unlabeled field seismic volumes via masked modeling. In parallel, a perception network is trained to extract high-level geological features—such as faults, stratigraphy, and geobodies—which are later used as frozen feature extractors to provide structure-aware supervision during fine-tuning via perceptual loss. In the supervised fine-tuning stage, GEM is optimized using heterogeneous prompt-label pairs and trained with adversarial and perceptual losses. These losses encourage geological plausibility across outputs, while enabling generalization to unseen tasks, data modalities, and regions. For large-scale 3D deployment, GEM employs a purely convolutional architecture (Figure \ref{fig10}), ensuring efficient inference without compromising spatial coherence. The training data preparation is detailed in Appendices \ref{dataprep}, and the model architecture and training process are described in the Methods section.


\subsection{Emergent Structure-Aware Reasoning in GEM}
\begin{figure}[!h]
	\includegraphics[scale=0.245]{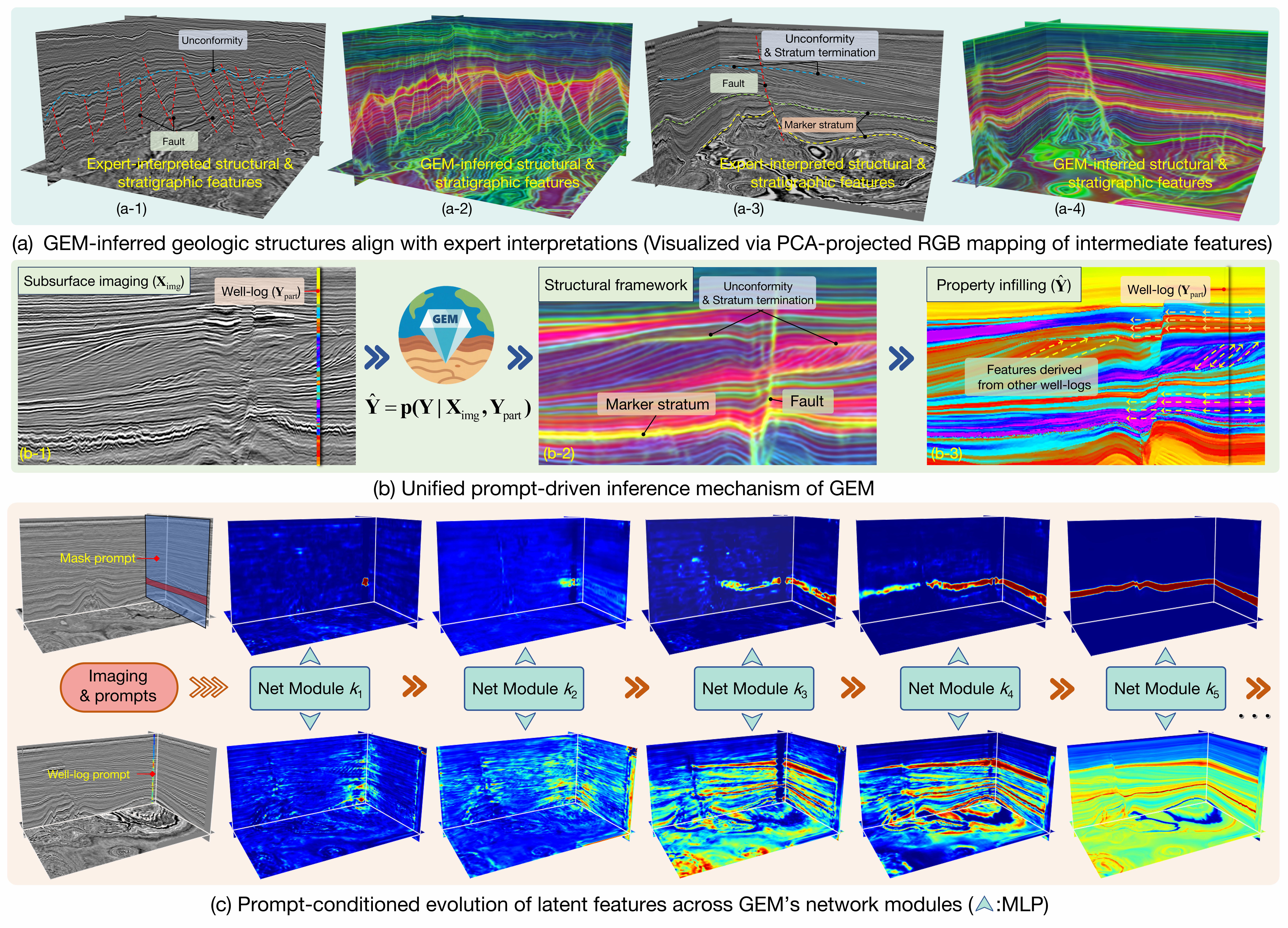}
	\centering\caption{
		Internal structure-aware reasoning mechanism of GEM. (a) Comparison between expert interpretations and GEM’s internal representations of seismic volumes. Dashed lines highlight how geologists identify key geological structures—such as faults, unconformities, and marker strata. Without using these annotations as inputs, GEM’s latent features capture similar patterns, revealing structure-aware reasoning consistent with expert understanding. This structure-awareness underpins GEM’s ability to generalize across diverse subsurface interpretation and modeling tasks. (b) Conceptual diagram of GEM’s unified inference mechanism. Given seismic input and human prompts (e.g., logs, masks, sketches) (left), GEM generates outputs (right) by reasoning along the inferred structural framework (middle). This formulation supports diverse subsurface imaging tasks within a single model. (c) Visualization of prompt-guided feature propagation. Different prompts guide latent feature evolution along geological structures, enabling geologically consistent outputs across task types.	}
	\label{fig05}
\end{figure}

The key to GEM’s superior performance and generalization ability lies not only in prompt-based flexibility, but also in its emergent capacity to recognize and reason upon subsurface geological frameworks—much like a human interpreter. Rather than approaching subsurface tasks as isolated segmentation or modeling problems, GEM formulates them as processes of identifying and completing a latent geological framework inferred from subsurface images. This mirrors expert geological reasoning, which emphasizes spatial continuity, stratigraphic hierarchy, and structural disruptions such as faults and unconformities. Through extensive self-supervised learning and fine-tuning with heterogeneous supervision, GEM develops an internal structural awareness that enables it to recognize key geological features directly from subsurface imagery. This capability grounds GEM’s ability to generate geologically plausible outputs across diverse settings and tasks, forming the foundation of its zero-shot adaptability.

Figure \ref{fig05} (a) illustrates this emergent structural understanding of subsurface during inference. To visualize GEM’s internal representations computed from input seismic volumes (a-1, a-3), we extract feature maps from the first upsampling stage and apply Principal Component Analysis (PCA) along the channel dimension to reduce them to three principal components. These reduced features are then projected into the RGB color space, resulting in visualizations (a-2, a-4) that consistently and coherently highlight critical geological elements such as faults, unconformities, and marker horizons. The spatial patterns of these features closely align with expert interpretations, as indicated by the dashed curves in (a-1, a-3). Note that the expert annotations are not used as inputs or prompts. This observation confirms that GEM can infer high-level geological concepts through its internal representations, without any explicit structural guidance during inference. 

Building on this internal representation, GEM performs prompt-conditioned inference by propagating sparse user inputs—such as well logs, masks, or sketches—along the inferred geological framework. As illustrated in Figure \ref{fig05} (b), one example involves using sparse well-log measurements as prompts (colored vertical trace in (b-1)) for physical property modeling. Because GEM can internally infer a geologic framework from the seismic image (b-1)—including features such as faults, unconformities, and marker strata—it can propagate these well-log values in a geologically consistent manner. This enables the generation of property volumes that honor both the structural context of the subsurface image and the sparse well-log conditioning. This process demonstrates how GEM leverages its internal structure awareness to produce geologically plausible outputs from minimal conditioning data.

To better understand how prompts guide internal reasoning, we freeze the pretrained GEM model and attach a lightweight multilayer perceptron (MLP) to probe its intermediate features during forward inference. As shown in Figure \ref{fig05} (c), we examine how different prompt types (a 2D mask or a vertical well-log) trigger distinct patterns of feature activation. These MLP-projected visualizations reveal how prompt information propagates through successive network modules ($k_1$ to $k_5$), gradually expanding from local prompt regions into structurally coherent representations. The mask prompt activates localized features that spread laterally along a geologic stratum, while the well-log prompt initiates vertical propagation that aligns with stratigraphic continuity. This behavior mirrors expert interpretation: starting from sparse observations, GEM progressively extrapolates geologically consistent structures that respect boundaries and relationships encoded in the subsurface image. Such prompt-aware, structure-conditioned reasoning underlies GEM's ability to unify diverse interpretation tasks---ranging from structural delineation and geobody segmentation to physical property modeling---within a single, generalizable framework.

Together, these visualization results reveal that GEM’s internal structure perception is not merely a byproduct, but a prerequisite for generalizable, zero-shot inference, and a core enabler of unified subsurface interpretation across diverse geoscientific tasks. In the following sections, we demonstrate how this capability supports high-quality results across a wide range of subsurface imaging tasks.

\subsection{Structural and Stratigraphic Interpretation}
\begin{figure}[!h!t]
	\includegraphics[scale=0.525]{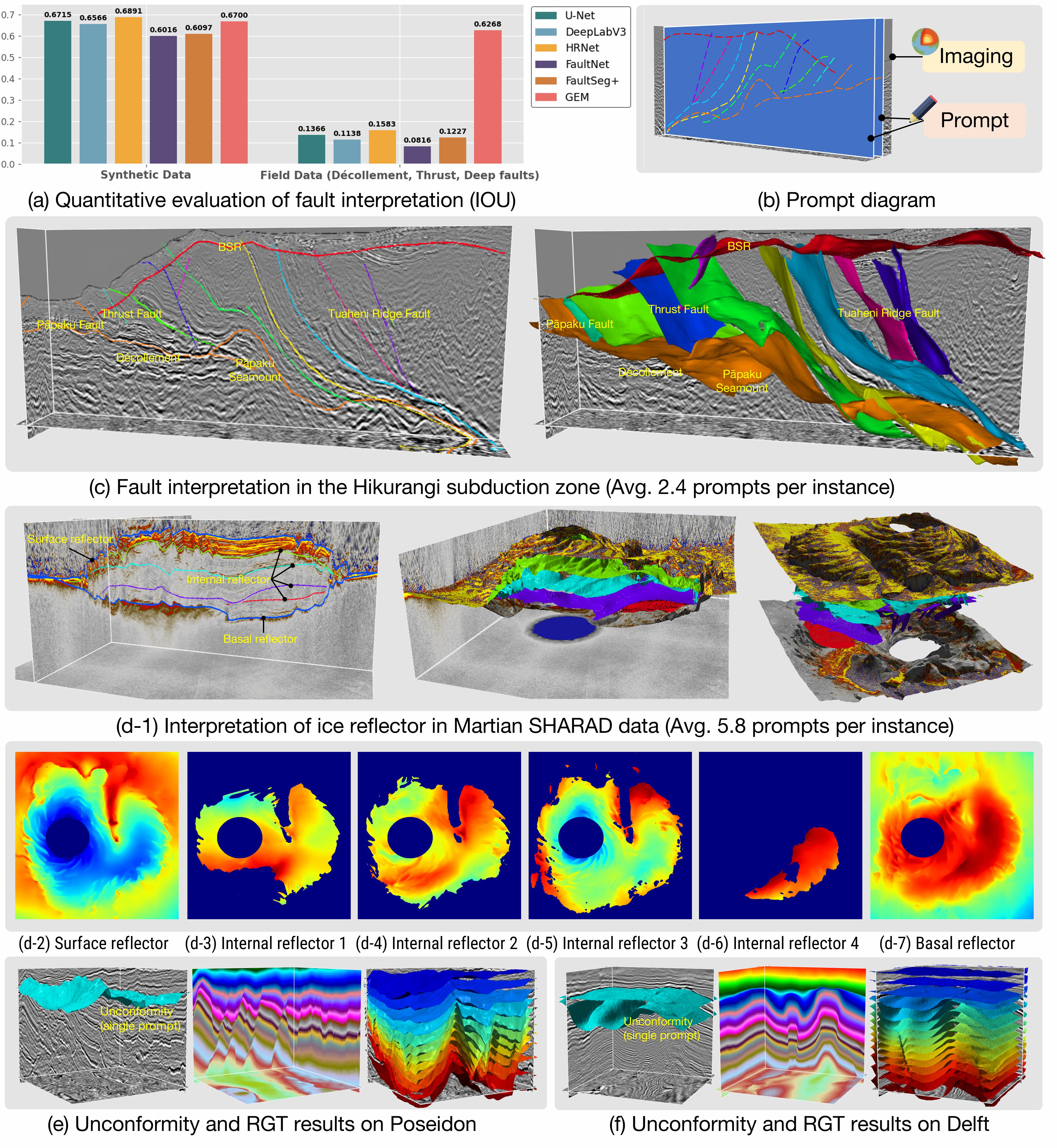}
	\centering\caption{GEM demonstrates strong zero-shot performance across diverse subsurface interpretation tasks. (a) Quantitative evaluation of fault interpretation using intersection-over-union (IOU) on both synthetic and field datasets. GEM consistently outperforms baseline models across multiple fault types, without task-specific retraining. (b) Schematic illustrating how expert prompts (e.g., BSR and fault sketches) are combined with seismic imaging to guide GEM’s inference. (c) Zero-shot 3D interpretation of the Hikurangi subduction margin. With an average of 2 prompts per instance, GEM reconstructs a full-volume BSR surface and complex thrust fault network. (d) Interpretation of Martian polar stratigraphy from SHARAD radargrams. Panel (d-1) shows input with prompt location; panels (d-2) to (d-7) visualize surface and internal reflectors generated by GEM. The model demonstrates emergent generalization to an unseen radar modality using only a few expert-provided mask prompts. (e, f). Unconformity and RGT estimation on the Poseidon and Delft datasets. GEM produces geologically consistent RGT volumes and boundary surfaces from sparse initial constraints.  More results are presented in Figure \ref{fig07}.}
	\label{fig02}
\end{figure}

Interpreting subsurface structures and stratigraphy remains fundamentally constrained by the highly heterogeneous and non-stationary nature of sedimentary systems. These complexities arise not only from primary depositional heterogeneities but are further compounded by post-depositional geological processes such as faulting, folding, karstification, magmatic intrusions, salt tectonics, and diagenesis. Such processes disrupt original stratigraphic architectures through  deformation dislocation and truncation leading to disordered and discontinuous sequences that deviate significantly from idealized layer-cake models. Subsurface imaging techniques, such as seismic imaging, are further limited by acquisition bandwidth, resolution, signal-to-noise ratio, and inaccuracies in velocity modeling—factors that reduce imaging fidelity, and semantic richness, ultimately contributing to significant interpretational ambiguity.

A representative class of such ambiguous structures is low-angle faults, including décollements and thrusts, whose seismic reflectivity often closely mimics that of stratigraphic interfaces or horizons. In the absence of geological priors, AI models struggle to distinguish these features from continuous stratigraphy. Once trained on such ambiguous patterns, models tend to overgeneralize, misclassifying horizons as faults—a learned bias driven by limited contextual cues. This phenomenon highlights a deeper epistemic tension: the mismatch between the low semantic density and resolution of subsurface data and the high structural complexity of the geological phenomena they encode. It represents a central bottleneck in the deployment of AI for seismic interpretation, where feature ambiguity, geologic non-uniqueness, and weak supervision hinder robust pattern discrimination.

GEM addresses this challenge by serving as a human-in-the-loop geophysical foundation model that enables expert-guided interpretation across diverse geological settings. It allows interpreters to interactively compute fault surfaces, horizons, or RGT volumes through multimodal prompts. We validate GEM across a diverse set of challenges, including décollement and thrust fault interpretation in the Hikurangi margin \cite{bangs2023slow}, bottom simulation reflector (BSR) tracking, Martian ice reflector extraction using SHARAD radargrams \cite{foss20173d}, unconformity surface extraction, and RGT volume estimation or fully stratigraphic interpretation in the Poseidon and Delft surveys  \cite{OpendTect}. Additional case studies are provided in the Appendices (Figure \ref{fig07}), demonstrating GEM’s versatility in both terrestrial and extraterrestrial datasets.

\subsubsection{Quantitative evaluation metric}

Before presenting these case studies in detail, we first present GEM’s overall quantitative performance on both synthetic and field datasets. The synthetic dataset consists of 100 volumes, of size $512\times512\times512$. For field dataset includes 20 expert-annotated slices from the Hikurangi Margin and Baiyun surveys (Figure \ref{fig07} (b)).

We compared GEM with three widely used AI models: UNet \cite{ronneberger2015u}, DeepLabV3 \cite{chen2018encoder}, and HRNet \cite{9052469}. To ensure fairness, the same prompt inputs provided to GEM were incorporated as weak labels into the training sets of these baseline models. Additionally, we evaluated two open-source fault detection models, FaultNet \cite{dou2022md} and FaultSeg plus \cite{wu2019faultseg3d, li2024faultseg3d}, both of which have achieved state-of-the-art performance on multiple datasets. These models were included for both quantitative comparison and qualitative result analysis.

As shown in Figure \ref{fig02} (a), GEM does not exhibit a significant advantage on synthetic data, with only minor performance differences compared to conventional AI models. This is primarily due to the high consistency between the features and distributions of the synthetic testing and  the training data, which allows traditional models to effectively learn underlying statistical patterns and inductive biases.
However, in tests on field data, GEM demonstrates clear superiority.  Under the same synthetic training conditions, GEM enables zero-shot complex geological structures through user prompts, significantly outperforming conventional models that rely on static training  (as shown in the right side of Figure \ref{fig02} (a)).

\subsubsection{Fault interpretation of the Hikurangi subduction zone}
We focus on the task of identifying and characterizing specialized, complex fault structures along the Hikurangi margin\cite{bangs2023slow}, a region that poses substantial challenges for automated interpretation. This subduction zone margin is heavily influenced by collisions with rough basement associated with seamounts on the subduction plates. The relief from subducting seamounts introduces complex, non-planar fault geometries, heterogeneous lithologies created by the influence of seamounts on basin deposition, and heterogeneous sediment compaction associated with stress shadows created by seamounts. These features significantly complicate fault interpretation, particularly for décollements, thrust systems, and fluid-rich sedimentary lenses associated with seamount induced stress shadows. The intricate spatial relationships between structures, combined with subtle seismic expressions, often hinder reliable detection using more conventional automated techniques.

Existing methods, such as FaultNet and FaultSeg+, have been widely adopted in industry practice. However, as illustrated in Appendices Figure \ref{fig07} (a), these models struggle in this challenging subduction zone setting—they tend to overfit to secondary normal faults and frequently miss the key structural elements that control slip behavior along the margin. Consequently, practitioners often resort to manual interpretation, which entails tracing faults across numerous 2D seismic sections followed by interpolation into 3D surfaces. This process can take weeks to complete\cite{bangs2023slow}, and it often lacks geometric precision and fails to capture fine-scale fault morphology critical for mechanistic analysis.

In contrast, GEM achieves significantly improved results with minimal user input. As shown in Figure \ref{fig02} (b, c), GEM requires only 2.4 user interactions per 3D fault surface, on average, to construct a complete and geologically consistent fault network. Despite being trained solely on synthetic data featuring normal faults, GEM generalizes effectively to identify complex fault types including décollements, thrusts, and deep-seated (see Figure  \ref{fig07} (b) Baiyun survey) structures. Its predictions align closely with expert interpretations and preserve structural details that are often lost in conventional methods, enabling instance-level precision in 3D fault surface delineation.

These enhanced results provide crucial support for interpreting the mechanical framework of the Hikurangi margin. For instance, the data reveals that the Pāpaku fault—developing along the trailing edge of a subducting seamount—forms a fluid-rich sedimentary lens in the stress shadow of the seamount that maintains high pore pressure and impedes overlying plate collapse. Similarly, the spatial coincidence between ancient larger, earlier formed sedimentary lens beneath the Tuaheni fault system and the 2014 slow slip zone suggests that such under-compacted sediment lenses exert sustained control on slow slip activity. GEM facilitates the extraction and 3D reconstruction of these complex fault systems with high precision, enabling geoscientists to more efficiently and accurately delineate the fault geometries that underpin these observations. By reducing reliance on time-consuming manual interpretation, GEM accelerates the process of linking seismic structure to geomechanical behavior, thereby enhancing our ability to interpret the role of fault architecture and fluid conditions in controlling shallow slow slip events.

\subsubsection{Emergent structural interpretation of unseen Martian SHARAD radagrams}
The detection of faults in the Hikurangi margin demonstrates GEM's strong capability in seismic structural  interpretation, achieving instance-level delineation of multiple fault types that remain challenging for existing AI methods. Beyond seismic imaging, GEM also generalizes to entirely new types of subsurface data that were not included in its training set. A notable example is its application to SHARAD radargrams acquired by the Mars Reconnaissance Orbiter (MRO) \cite{foss20173d}, where dense orbital coverage enables 3D imaging of the polar layered deposits (PLDs) within Mars’ polar ice caps.

We evaluated GEM’s capability to interpret complex internal reflectors within the Martian polar ice caps using 3D SHARAD radar data. These 3D datasets, derived from thousands of orbital radar soundings, reveal the intricate stratigraphy of Planum Boreum and Planum Australe, which record layered deposition, unconformities, and possible impact-related structures accumulated over millions of years of climate cycling on Mars. However, interpreting these internal layers is highly challenging due to limited vertical resolution, low signal-to-noise ratio, and frequent clutter arising from surface topography and ionospheric distortions. Critically, there are currently no effective automated methods for extracting internal radar reflectors, and manual interpretation remains the norm—an inefficient and subjective process that limits scientific inference.

Remarkably, despite being trained solely on synthetic seismic data with no exposure to radar or ice-layer labels, GEM successfully generalizes to unseen SHARAD radargrams, which differ fundamentally from seismic data in their underlying imaging physics. As shown in Figure \ref{fig02} (d), GEM accurately identifies  several key reflectors in the SHARAD dataset, including the ice surface, the basal reflector, and four internal reflectors. The ice surface marks the boundary between the ice body and the overlying atmosphere or debris layer, while the basal reflector defines the interface between the ice and underlying bedrock or sediment. These picked 3D reflector surfaces provide valuable insight into the internal architecture and evolutionary history of  Martian ice deposits. Analyzing the geometry, spatial extent, and relationships among the surface, basal, and internal reflectors helps reveal depositional cyclicity, ice flow dynamics, and records of intermittent melting and refreezing. These findings provide direct evidence for the stratification of Martian paleoclimate and the mechanisms underlying climate transitions. Moreover, the identified reflector architecture offers practical benefits—enabling assessment of ice purity and mechanical stability, which are essential for planning future drilling operations and in situ resource utilization.

GEM’s ability to interpret layered radar structures—despite being trained solely on seismic volumes—highlights its emergent understanding of generic subsurface structural patterns. Although the physical principles of radar and seismic imaging differ, both encode material contrasts through reflectivity. GEM appears to have learned high-level priors—such as interface continuity, reflector geometry, and hierarchical layering—that generalize across modalities. This emergent generalization suggests the model is not simply fitting data-specific features but instead internalizing abstract representations of subsurface structure. Such cross-modal adaptability, acquired without explicit radar training, points to the potential of foundation models like GEM to support universal subsurface interpretation across diverse geological settings, sensor types, and planetary environments.

\subsubsection{Fully stratigraphic interpretation}

In addition to structural interpretation, GEM also generalizes well to interpret unconformities (cyan surfaces in Figure \ref{fig02} (e, f)) and estimate a RGT volume for achieving fully stratigraphic interpretation\cite{bi2021deep,yang2023multi}. 
By constructing a stratigraphic  time field that represents isochronous surfaces throughout the seismic volume, RGT captures detailed stratification, unconformities, and complex structural features, while effectively encoding information about depositional sequences and geological structures. Compared with traditional discrete horizon picking methods, RGT offers a dense, coherent, and geologically consistent representation, serving as a unified foundation for tasks such as stratigraphic interpretation, sedimentary analysis.

RGT estimation remains a highly challenging task, particularly in the presence of unconformities and fault systems that disrupt stratigraphic continuity and introduce structural non-stationarity. Traditional methods often require extensive manual interpretation and struggle to maintain global consistency across such complex geological features.
With GEM, RGT estimation becomes a prompt-driven process. As shown in the Appendices Figure \ref{fig12}, we begin with a uniformly distributed initial RGT scale, serving as a rough timeline to guide inference. GEM then refines the RGT field through several forward passes, progressively adjusting it to match stratigraphic and structural features in the data (see Appendices \ref{rgtapdx} for details).
We evaluate GEM's RGT estimation on two field datasets of Poseidon and Delft.

The Poseidon dataset is located in the Browse Basin on the northwestern shelf of Australia, a region marked by complex geological structures including dense normal fault networks and prominent unconformities. The Delft dataset originates from the West Netherlands Basin in the western Netherlands, characterized by reverse faults, unconformities, and igneous intrusions that reflect a complex tectono-sedimentary history.  Despite their distinct tectonic settings, both datasets exhibit severe structural complexities including dense fault systems, prominent unconformities, and igneous intrusions. These features lead to stratigraphic truncation, fault-induced displacement, and intrusive disruption which typically hinder reliable RGT estimation and horizon interpretation. Despite these complexities, GEM delivers geologically reasonable RGT volumes (center panels of Figure \ref{fig02} (e, f) ) that preserve stratigraphic consistency across faults while allowing for appropriate jumps or terminations  at unconformities and intrusive boundaries. By extracting iso-surfaces from the RGT volume, a complete set of stratigraphic horizon surfaces is reconstructed (right panels of Figure \ref{fig02} (e, f)).  The ability to estimate a geologically consistent RGT volume and extract all horizons across the entire seismic volume further validates GEM’s strong understanding of the global structural and stratigraphic framework, as well as its capability to capture fine-scale geological details embedded within complex 3D seismic data.

\subsection{Geobody Segmentation}
\begin{figure}[!h]
	\includegraphics[scale=0.40]{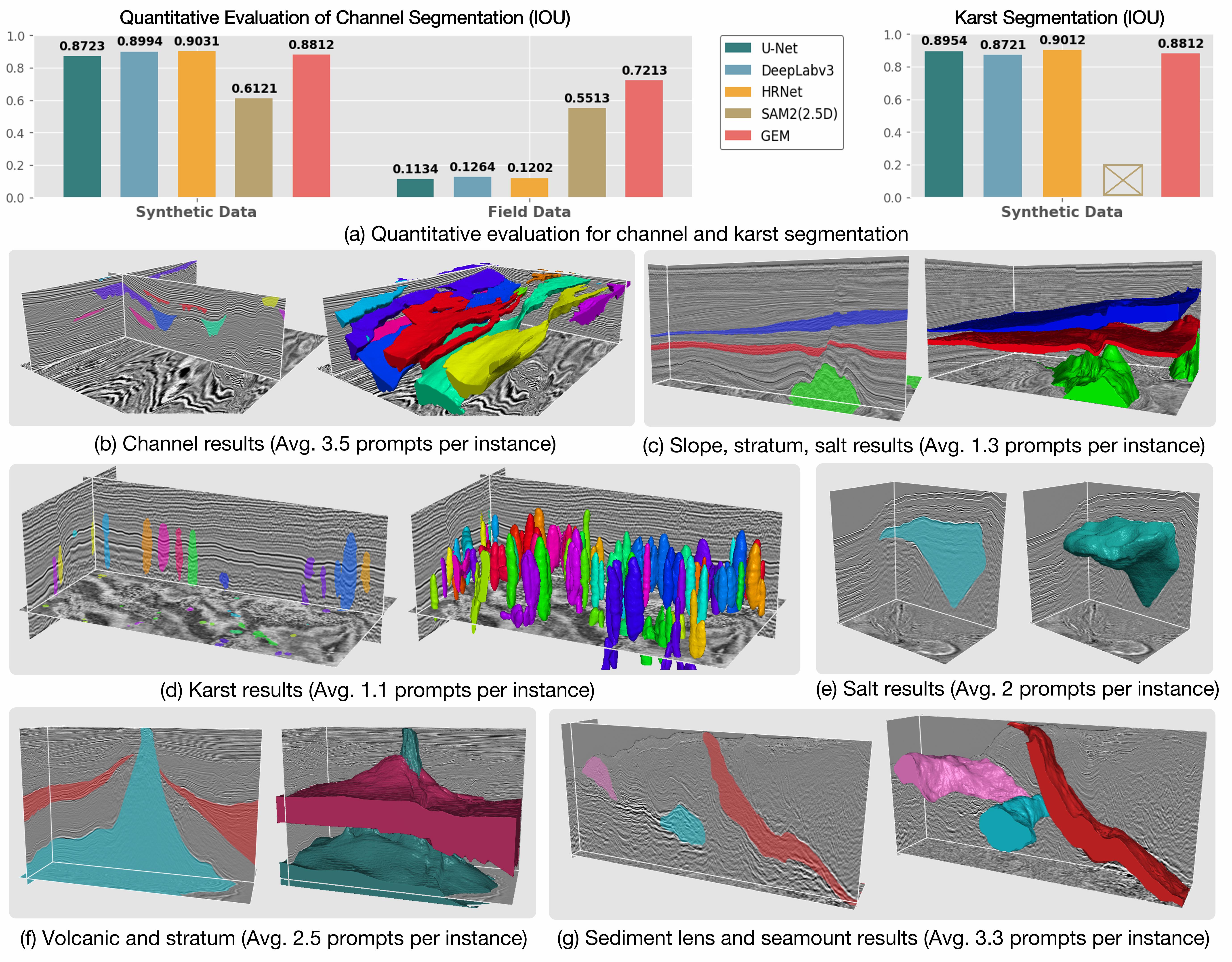}
	\centering\caption{GEM demonstrates generalizable geobody segmentation across diverse geological settings. (a) Quantitative comparison on synthetic channel and karst datasets. GEM significantly outperforms conventional models on real seismic data, demonstrating strong generalization without retraining. (b) Segmentation of complex paleo-channels in the Parihaka survey. Despite strong faulting and ambiguous reflections, GEM produces coherent 3D channel structures. (d) Karst systems in the Fort Worth Basin. (c, e, f, g) Zero-shot generalization to diverse geobodies not present in training, including slope and salt bodies in the Netherlands F3 (c), salt domes in a Gulf Coast dataset (e), volcanic units and stratums in the Kora region (f), and a subducting seamount with sediment lens in the Hikurangi margin (g). These results highlight GEM’s ability to produce geologically consistent, instance-level segmentations with minimal prompting. More results are presented in Figure \ref{fig08}.}
	\label{fig03}
\end{figure}

Geobody segmentation tasks currently face significant challenges in cross-survey generalization. As discussed in Appendices \ref{apd1}, this is primarily due to the semantic incompleteness of images obtained through existing subsurface imaging techniques. Without incorporating human expertise, it remains difficult to distinguish various geological bodies—such as salt bodies, channels, karsts, and volcanic units—based solely on texture, gradient, and morphological cues present in the data. In contrast, GEM enables zero-shot generalization to arbitrary geobodies under human guidance, and its performance consistently surpasses that of existing customized models across multiple field datasets.

The synthetic test set for channels was derived from the CIGChannel \cite{wang2024cigchannel} dataset with consists of 100 samples. The synthetic test set for karsts was taken from the KarstSeg \cite{wu2020deep} dataset, comprising 20 samples. For real data validation, 30 two-dimensional slices were randomly selected from the Parihaka and Romney (Figure \ref{fig08} (b-1)) datasets and manually labeled for channel evaluation. The baseline methods used for comparison experiments include UNet, DeepLabV3, and HRNet.

In the quantitative experiments on synthetic data, GEM does not exhibit a clear advantage (left side of Figure \ref{fig03} (a)), similar to the case in structural interpretation tasks. Its strengths lie in generalization to field data and its powerful zero-shot learning capabilities. Particularly in the task of channel segmentation, the diversity of fluvial depositional patterns and structural complexity sharply contradicts the low information density and semantic deficiency inherent in subsurface imaging. This contradiction is a major factor limiting the generalization ability of most existing channel detection methods across different survey areas, often resulting in a typical IOU below $0.2$ for conventional AI models. Their qualitative outputs are also unreliable, frequently missing key channel structures or introducing significant false positives, as illustrated in Appendices Figure \ref{fig08} (a, b).

\subsubsection{Channel segmentation}

Figure \ref{fig03} (b) shows the result for channels in the Parihaka dataset. Detecting paleo-channels in this region is particularly challenging due to their highly variable scale, complex morphology, and diverse seismic reflection characteristics, including high-amplitude zones, chaotic reflections, and weak boundary signals. Additionally, the area is structurally complex, with well-developed fault systems that introduce significant seismic discontinuities. These features often confuse AI models, which may misinterpret fault-induced discontinuities as channel-related erosion features.

Different depositional facies in this region exhibit similar seismic responses, further complicating feature discrimination. These challenges have made Parihaka a longstanding benchmark for evaluating channel detection methods \cite{wang2024cigchannel, gao2024foundation}. GEM successfully captures the spatial morphology of the paleo-channel within the 3D seismic volume and achieves instance-level segmentation. This result has the potential to reveal sedimentary processes and paleoenvironmental evolution since the Miocene, while also providing critical insights into reservoir geometry and connectivity analysis.

Moreover, quantitative results indicate that both GEM and SAM2 are capable of performing channel segmentation when guided by human-provided prompts (Figure \ref{fig03} (a)). Following the standard video segmentation pipeline of SAM2, we applied it to the Romney dataset, with results shown in Appendices Figure \ref{fig08}. Although SAM2 is capable of 3D segmentation, its backbone remains fundamentally two-dimensional. As a result, the segmentation exhibits pronounced jagged discontinuities along the crossline direction. In contrast, the outputs generated by GEM are smooth and coherent in all directions.

\subsubsection{Karst segmentation}

Figure \ref{fig03} (d) illustrates the application of GEM for karst geobody interpretation in the Fort Worth Basin, a hydrocarbon-rich foreland basin in North Central Texas. This region is geologically characterized by extensive paleokarst development within the Ellenburger carbonate formation, formed during prolonged subaerial exposure from the Late Ordovician to Early Pennsylvanian. Subsequent burial and compaction induced collapse structures, resulting in complex karst systems that appear as distinct geomorphological features in seismic data.

GEM achieves instance-level karst segmentation directly through minimal expert interaction and prompt guidance, without requiring any fine-tuning or prior training on the Fort Worth dataset. This highlights GEM's  advantage over conventional semantic segmentation models, enabling geologically meaningful delineation of complex subsurface features. As shown in Figure \ref{fig03} (d), GEM successfully reconstructs multiple discrete karst geobodies with clear morphological boundaries and stratigraphic context, demonstrating its potential for detailed structural characterization, reservoir modeling, and field development planning.

\subsubsection{Other geobody segmentation tasks}

GEM is not limited to segmenting channels and karst features included in its training dataset. Instead, it demonstrates strong generalization capabilities to a broad range of unseen subsurface geobodies, as illustrated in Figure \ref{fig03} (c, e, f, g).

Figure \ref{fig03} (g) showcases GEM’s segmentation of a sediment lens and a subducting seamount in the Hikurangi margin. The sediment lens, formed in the trailing wake of the seamount, retains undercompacted, fluid-enriched sediments, creating a localized zone of high pore pressure and low effective stress—key conditions that promote the occurrence of slow slip events. Accurately mapping the 3D geometry of such lenses is essential for identifying potential nucleation zones, assessing structural heterogeneity along the subduction interface, and improving forecasts of seismic and aseismic slip behavior.

The identification of seamounts further enhances understanding of subduction zone dynamics by revealing how these features regulate the regional stress field and deform surrounding faults. In the Hikurangi setting, the spatial relationships between the subducting seamount and faults such as Pāpaku, Tuaheni, and the décollement shed light on the mechanisms that form sediment lenses and influence fluid-driven slip. Moreover, the geometry and deformation of seamounts provide insights into long-term tectonic evolution and help locate potential slip concentration zones and stress accumulation regions—factors critical for earthquake and tsunami hazard assessment.

Notably, GEM accurately segments both the sediment lens and the seamount with an average of only 3.3 user interactions per geobody, despite not having been trained on either structure type. Similar zero-shot generalization is observed across other geobodies in different geological settings, including slope bodies, stratums, volcanic units, and salt bodies. These results collectively highlight GEM’s robustness, versatility, and strong adaptability to geologically diverse and structurally complex environments.

\subsection{Well-Log Prompted Geophysical Property Modeling}
\begin{figure}[!h]
	\includegraphics[scale=0.45]{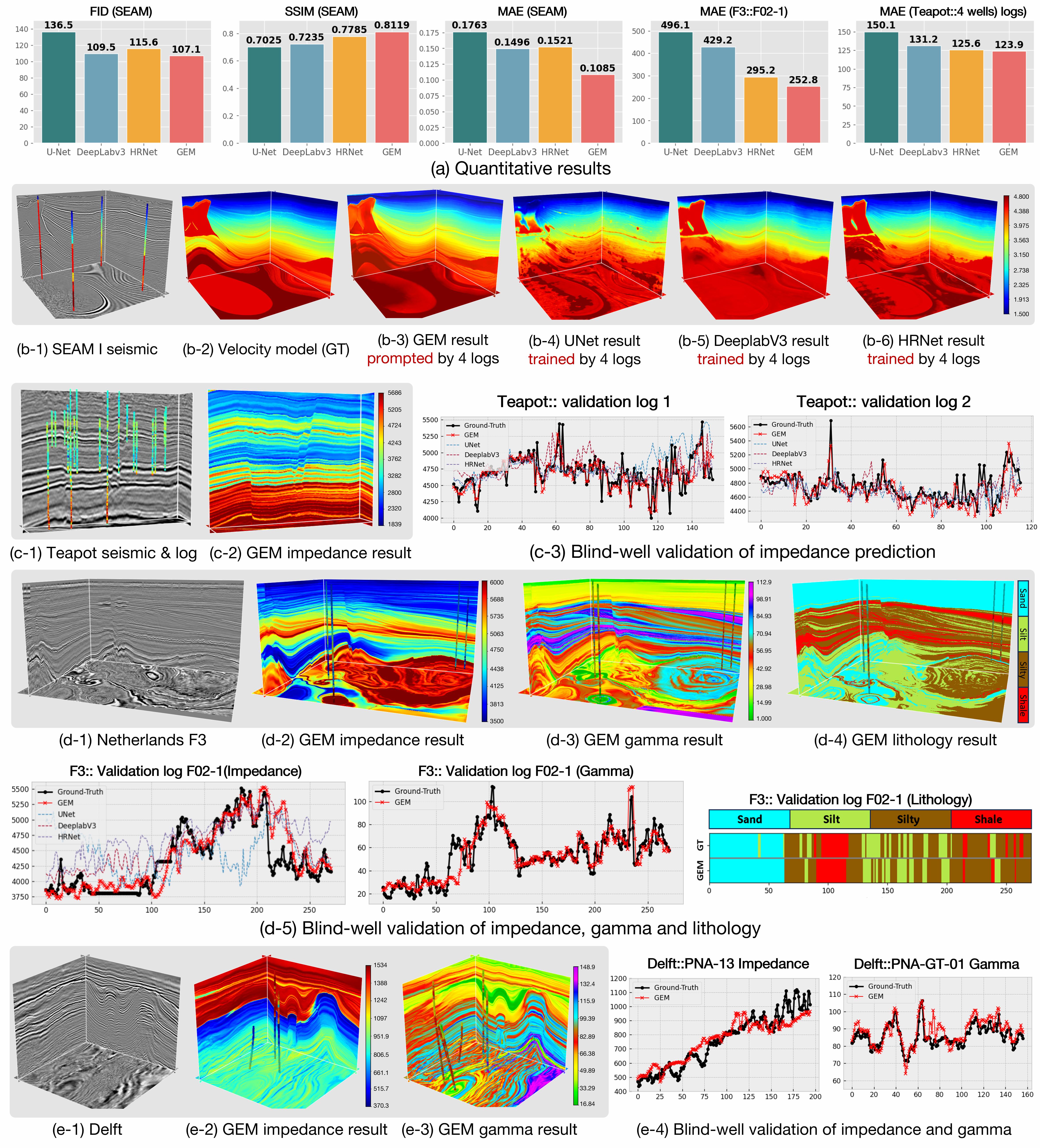}
	\centering\caption{GEM enables prompt-guided, multi-physical-property modeling with minimal well control. (a) Quantitative comparison of impedance and gamma ray modeling on synthetic (SEAM I) and field datasets (F3, Teapot). In zero-shot settings using only sparse well-log prompts, GEM outperforms specifically trained conventional models including U-Net, DeepLabv3, and HRNet. (b) Qualitative comparison on SEAM I. Panel (b-3) shows GEM’s result prompted by 4 logs; panels (b-4) to (b-6) show results from models trained with the same logs. (c) Impedance modeling (c-2) on the Teapot Dome dataset, with blind well validation shown in (c-3), where GEM better captures impedance variations with higher fidelity. (d) Multi-physical-property modeling on the Netherlands F3, including impedance (d-2), gamma ray (d-3), and lithology (d-4), based on sparse log inputs. Panel (d-5) shows blind well validation, where GEM’s predictions remain highly consistent with ground truth. (e) Using sparse log prompts in another survey of Delft (e-1), GEM generates geologically consistent impedance (e-2) and gamma (e-3) volumes with strong agreement at non-prompted wells (e-4). More results are provided in Figure \ref{fig09}.}
	\label{fig04}
\end{figure}

Accurately modeling geophysical properties such as velocity, acoustic impedance, gamma ray, and lithology from seismic and well-log data is essential for reservoir characterization and subsurface analysis. However, conventional learning-based inversion methods require retraining for each new survey area, as they rely on paired seismic–well data to capture region-specific property trends. This dependence severely limits generalization: seismic signals often lack low-frequency components needed to preserve large-scale trends, and well logs—when sparse—can easily lead to overfitting. Moreover, conventional models treat each property prediction task independently, ignoring the shared structural context that governs subsurface variability \cite{dou2024contrasinver,wu2021deep,wu2022seismic}.

GEM addresses these challenges by treating well logs as flexible prompts that introduce regional property trends, while relying on its pretrained ability to extract geological frameworks (e.g., structural boundaries and stratigraphic sequences) from the input seismic volume to guide physical property modeling. Instead of learning a direct mapping from data to property, GEM performs welllog-prompt-driven property completion constrained by the inferred geologic architecture.This enables the generation of property volumes that are not only geologically plausible, but also spatially consistent with seismic structures.
As a result, GEM supports zero-shot modeling of diverse geophysical properties—even those not included in training—without retraining. As shown in Figure \ref{fig04}, we demonstrate this across four representative datasets: SEAM I (velocity), Teapot Dome (impedance), Netherlands F3 (impedance, gamma ray, lithology), and Delft (impedance and gamma ray), where GEM consistently produces high-fidelity property models from minimal well input.

To demonstrate GEM’s modeling capabilities, we compare its performance against three conventional AI models trained under supervised, multidimensional inversion frameworks. While these baselines rely on paired seismic–well data and task-specific training, GEM performs inference directly using well
logs as prompts, without any retraining. In the quantitative results, for the synthetic dataset, where a complete 3D velocity volume is available, we report both semantic-level and patch-level performance metrics, including Frechet Inception Distance (FID) \cite{Seitzer2020FID} and Structural Similarity Index (SSIM) \cite{wang2004image}. For the field datasets, that lack full-volume ground truth, evaluation is conducted using pixel-level Mean Absolute Error (MAE) calculated at blind well locations—that is, wells excluded from both the training set of conventional models and the prompt input to GEM. This ensures a fair and consistent comparison of generalization performance across methods. The quantitative metrics in Figure \ref{fig04}a clearly demonstrate GEM’s superior performance in physical property modeling: it achieves consistently higher SSIM and lower MAE on synthetic data, and lower MAE on field datasets, outperforming conventional supervised models across all test cases.

\subsubsection{Velocity modeling on the SEAM I}

The SEAM I dataset provides a synthetic yet structurally complex benchmark for velocity modeling, featuring high-contrast salt bodies, lateral heterogeneity, and interlayer variability. These characteristics pose strong challenges for data-driven inversion, especially under sparse well control.

Figure \ref{fig04} (b) shows the ground-truth velocity model, while Figure \ref{fig04} (b-3) presents the velocity model generated by GEM using only four wells. GEM successfully reconstructs the typical salt-related structures in the SEAM I dataset, including salt bodies and salt tongues. It also accurately captures the complex velocity gradient from the shallow sedimentary layers to the deep basement, preserving both high-resolution interlayer velocity variations and strong lateral heterogeneity.

In contrast, the baseline methods, trained on the same spare well logs, suffer from severe overfitting due to the limited training data. Although DeeplabV3 (Figure \ref{fig04} (b-5)) and HRNet (Figure \ref{fig04} (b-6)) manage to partially reconstruct the salt geometries, their results are generally affected by noise and exhibit lower spatial resolution, making it difficult to accurately capture stratigraphic details and velocity transitions.

\subsubsection{Acoustic impedance modeling on the Teapot Dome}
The Teapot Dome survey is an open-access land-based seismic survey provided by the U.S. Department of Energy (DOE) and the Rocky Mountain Oilfield Testing Center (RMOTC). It covers the Teapot Dome structural region in Natrona County, Wyoming. The geological structure of this dataset is relatively simple, and it includes extensive acoustic impedance well logs, making it well suited for evaluating and benchmarking inversion algorithms. This dataset lies in the frequent alternation of thin sandstone and mudstone layers, with thicknesses approaching the vertical resolution of the seismic data. These layers exhibit low impedance contrast and weak reflection responses, making accurate inversion of thin-layer structures particularly difficult. This imposes high demands on the resolution capacity of inversion methods, their ability to recover low-frequency content, and their robustness under the non-uniqueness constraints imposed by limited seismic bandwidth.

In Figure \ref{fig04} (c), GEM demonstrates superior performance in capturing thinlayer structures with greater clarity and continuity. It effectively suppresses the vertical ringing and lateral instability commonly observed in traditional methods (Figure \ref{fig09}). While maintaining consistency with the seismic structural framework, the impedance volumes generated by GEM exhibit higher spatial resolution and stronger stratigraphic continuity, particularly in zones with thin interbedded sandstone and mudstone. Figure \ref{fig04} (c-3) presents the blind-well test results, demonstrating a high degree of alignment between GEM’s predictions and the impedance trends observed in well logs. The corresponding MAE values further confirm the model’s accuracy. In addition, Appendices Figure \ref{fig09} (a) provides a qualitative comparison of physical property modeling results between GEM and several baseline methods. GEM consistently outperforms the others, particularly in terms of vertical resolution and the preservation of geological structures.

\subsubsection{Multi-physical-property modeling on the Netherlands F3}

The Netherlands F3 dataset is a widely used benchmark for 3D seismic interpretation and physical property modeling. Geologically, it features multi-phase fault systems, progradational slope deposits, and a regionally extensive angular unconformity that separates genetically distinct stratigraphic packages. These structures present strong phase shifts, impedance discontinuities, and highamplitude reflectors—making physical property modeling in this dataset particularly challenging. Such complex features demand models capable of capturing nonlinear responses and maintaining stratigraphic coherence across discontinuities.

Compounding these challenges is the fact that only 4 wells are available for the entire survey area of approximately 24 km². This extreme sparsity of well control severely limits the performance of conventional data-driven approaches, which often require dense supervision to generalize effectively. Under such conditions, traditional inversion models tend to produce blurred transitions, spatial drift, or anomalous outputs—an issue clearly illustrated in Appendices Figure \ref{fig09} (b-3, b-4, b-5).

Figure \ref{fig04} (d) shows that GEM, using only three wells as prompts, accurately reconstructs key impedance structures across the full volume. The model maintains stratigraphic and structural continuity even across faults and unconformities. Blind-well validation at the fourth well confirms that GEM effectively captures both large-scale trends and fine-scale variations.
Moreover, GEM successfully models gamma ray and lithology volumes (Figure \ref{fig04} (d-4)), although neither property was used during training. The resulting 3D property models exhibit clear alignment with geological structure and depositional sequences, preserving layer continuity and fault displacements in a geologically consistent manner. In addition, the validation curves at the blind well also demonstrate excellent agreement with ground truth measurements across all predicted properties as shown in Figure \ref{fig04} (d-5). This example highlights GEM’s strong generalization ability and zero-shot adaptability in multi-property prediction.

\subsubsection{Multi-physical-property modeling on the Delft}

The Delft dataset presents a particularly challenging scenario for physical property modeling due to its complex depositional and tectonic history. The region is dominated by fluvial facies deposits with strong lateral heterogeneity, and is further complicated by igneous intrusion from depth, which have deformed overlying strata through structural folding and reverse faulting. These features disrupt stratigraphic continuity and create irregular distributions of physical properties, making it difficult for conventional methods to generate models that are both geologically plausible and spatially coherent.

As shown in Figure \ref{fig04} (e), GEM generates high-quality 3D models of both impedance and gamma ray, capturing major structural transitions and maintaining stratigraphic consistency throughout the volume. The predicted property distributions exhibit strong alignment with the structural and stratigraphic framework indicated in the seismic volume. Notably, these results are achieved using only two wells as prompts for impedance and five wells for gamma ray, respectively. Blind-well validation further confirms the reliability of these predictions. For impedance, the model accurately reproduces the overall trend at the blind well, with strong agreement in both amplitude and layering throughout the interval. For gamma ray, the predicted variation closely matches the true curve in the upper section, while some deviation is observed at depth—likely due to extrapolation uncertainty and limited low-frequency trend information in the seismic response. Despite these local discrepancies, the overall consistency with ground truth highlights GEM’s capacity for structure-aware generalization under minimal prompting.

\section{Method}
\subsection{Unified Perspective and Overall Architecture}

We propose a unified view in which subsurface structural and stratigraphic interpretation, geobody segmentation, and physical property modeling can treated as tasks operating at different levels of abstraction of a common geologic framework. Based on this insight, we designed GEM to unify these diverse tasks through a prompt-driven generative process that propagates interactive sparse prompts along structural features inferred from subsurface imaging data. By leveraging shared structural constraints—including stratigraphic continuity, fault systems, regional geobody characteristics, stratigraphic hierarchy and spatial context relationships—the model enables coherent task integration and collaborative optimization within a single framework.

To formalize this unified perspective, all related tasks are modeled as a conditional generative process:

\begin{equation}
	\hat{\mathbf{Y}} = \mathbf{p}(\mathbf{Y} \mid \mathbf{X}_{\text{img}}, \mathbf{Y}_{\text{part}})
\end{equation}
where $\mathbf{X}_{\text{img}}$ represents the subsurface imaging data that encodes the structural framework, and $\mathbf{Y}_{\text{part}}$ is the partial prompt representing interactive, human-provided sparse inputs—such as well-log properties, geobody masks, and fault or horizon segments—which both specify the intended task and encode human-prior structural constraints to guide the generation process. 
This formulation enables the model to extrapolate from sparse inputs ($\textbf{Y}_{\text{part}}$) and generate geologically plausible outputs ($\hat{\textbf{Y}}$) consistent with the underlying structural framework encoded in the subsurface image $\textbf{X}_{\text{img}}$.
The model is thereby encouraged to perform geologically meaningful extension and infilling conditioned on structural priors.

To realize this unified modeling strategy, GEM employs a two-stage training framework consisting of self-supervised pretraining followed by  supervised fine-tuning with mixed prompts and labels drawn from diverse tasks (Figure \ref{fig01} (b, c)).
The pretraining stage employs a masked image modeling strategy to learn generalizable structural representations from large-scale real and synthetic seismic data
This stage establishes a foundational feature representation for downstream tasks by learning the broad geological characteristics present in real data, thereby accelerating the convergence of subsequent task-specific training.
The fine-tuning stage supervises conditional generation with a mixture of task-specific prompts and labels from diverse interpretation tasks, implemented under a Generative Adversarial Network (GAN) framework \cite{goodfellow2020generative}. 
Together these design choices instantiate a physics informed deep learning paradigm in which geological constraints are injected as differentiable, structure aware guidance.

\subsection{Self Supervised Pre-training}
We adopt a BERT-style pretraining strategy for convolutional networks, specifically leveraging the Sparse Masked Modeling (SparK) approach \cite{tiandesigning}, with tailored modifications for the GEM backbone. During the masked image modeling (MIM) process in 3D data, we employ a higher masking ratio of 80\%, exceeding the 60\% used in the original SparK method. This adjustment is motivated by the inherent characteristics of seismic data, where neighboring slices exhibit high similarity and repeated textural patterns. Therefore, under the same missing ratio, reconstructing 3D seismic data is easier than reconstructing 2D images. Increasing the reconstruction difficulty encourages the model to develop more generalizable representations, thereby improving performance in downstream tasks.

The training dataset comprises a large collection of both field-acquired and synthetic seismic volumes. These volumes are randomly cropped into cubes of size $160\times 160\times160$. Pretraining is conducted on $8\times$ NVIDIA H20 GPUs with a batch size of 64, using the AdamW optimizer. The learning rate is set to $0.001$, \(\beta_1 = 0.9\), \(\beta_2 = 0.999\), and the training proceeds for 500,000 iterations.

\subsection{Structure-aware Perceptual Network}
Most existing perceptual networks are built upon two-dimensional natural images \cite{zhang2018unreasonable}, exhibiting limited performance when applied to 3D data and struggling to extract and enhance critical structural and geological features in seismic volumes. To overcome these limitations, we trained a structure-aware perceptual network specifically designed for seismic data, aiming to improve the accuracy of subsequent supervised tasks such as modeling and RGT estimation, as well as to strengthen the network’s representational capacity for seismic structures. The architecture of the structure-aware perceptual network is identical to that of GEM (Figure \ref{fig10}), with the only difference being that the base width $c$ is set to 24.

This network is trained on synthetic data, with input comprising seismic volumes, property volumes, and RGT. Supervision is provided through a segmentation task involving three categories of geological structures present in the synthetic dataset: faults, channels, and karst cavities. A hybrid loss function combining multiclass Dice loss and cross-entropy loss is employed, defined as,
\begin{equation}
	\mathcal{L}_{\text{seg}} =  \frac{1}{C} \sum_{c=1}^{C} \left( 1 - \frac{2 \sum_i p_{i}^{(c)} t_{i}^{(c)} + \epsilon}{\sum_i p_{i}^{(c)} + \sum_i t_{i}^{(c)} + \epsilon} \right)  -   \frac{1}{N} \sum_{i=1}^{N} \sum_{c=1}^{C} t_{i}^{(c)} \log(p_{i}^{(c)})
\end{equation}
where \( C \) is the number of classes, \( N \) is the number of spatial positions per sample (e.g., pixels or voxels), \( p_{i}^{(c)} \) and \( t_{i}^{(c)} \) denote the predicted probability and the one-hot ground truth for class \( c \) at position \( i \), \( \epsilon \) is a small constant for numerical stability.

Training of the perceptual network is conducted on $8\times$ NVIDIA H20 GPUs with a batch size of $96$, using the AdamW optimizer with a learning rate of $0.001$, \(\beta_1 = 0.9\), \(\beta_2 = 0.999\), and a total of 300,000 iterations.

\subsection{Supervised Fine-tuning with Mixed Prompting}

Due to the ambiguity and anisotropy inherent in subsurface imaging data, which often result in non-uniqueness, practical applications tend to prioritize the reliability of a solution among all possible outcomes rather than its absolute correctness. This consideration is particularly critical in tasks such as structural modeling, RGT estimation, and physical property modeling, where the objective is to recover geologically plausible structures from incomplete information. Under such circumstances, only relative correctness can be meaningfully defined.
To address this challenge, we adopt a relativistic GAN strategy \cite{jolicoeur2018relativistic}. This approach shifts the modeling objective away from generating a single correct solution and instead encourages the model to learn how to select relatively better and geologically more credible structures from among multiple plausible solutions. This modeling paradigm aligns more naturally with the inherent uncertainty and non-uniqueness of seismic data. The corresponding formulation is given as follows,

\begin{equation}
	\mathcal{L}_{\text{dis}} = 
	\mathbb{E}_{\substack{
			\mathbf{Y}_{\text{r}} \sim \textbf{p}_{\text{data}}(\mathbf{Y} \mid \mathbf{X}_{\text{img}}, \mathbf{Y}_{\text{part}}) \\
			\mathbf{Y}_{\text{f}} = G(\mathbf{X}_{\text{img}}, \mathbf{Y}_{\text{part}})
	}} 
	\left[
	\left(C(\mathbf{Y}_{\text{r}}) - \mathbb{E}[C(\mathbf{Y}_{\text{f}})] - 1\right)^2 +
	\left(C(\mathbf{Y}_{\text{f}}) - \mathbb{E}[C(\mathbf{Y}_{\text{r}})] + 1\right)^2
	\right]
\end{equation}
\begin{equation}
	\mathcal{L}_{\text{adv}} = 
	\mathbb{E}_{\substack{
			\mathbf{Y}_{\text{r}} \sim \textbf{p}_{\text{data}}(\mathbf{Y} \mid \mathbf{X}_{\text{img}}, \mathbf{Y}_{\text{part}}) \\
			\mathbf{Y}_{\text{f}} = G(\mathbf{X}_{\text{img}}, \mathbf{Y}_{\text{part}})
	}} 
	\left[
	\left(C(\mathbf{Y}_{\text{f}}) - \mathbb{E}[C(\mathbf{Y}_{\text{r}})] - 1\right)^2 +
	\left(C(\mathbf{Y}_{\text{r}}) - \mathbb{E}[C(\mathbf{Y}_{\text{f}})] + 1\right)^2
	\right]
\end{equation}
where \( G(\mathbf{X}_{\text{img}}, \mathbf{Y}_{\text{part}}) \) denotes the conditional generator, which produces candidate structures and \( C(\cdot) \) represents the unnormalized confidence score output by the discriminator. During training, a subset of the 3D label volume $\textbf{Y}$ is randomly selected as a prompt ($\textbf{Y}_{\text{part}}$). This prompt can take arbitrary shapes that are easily interpretable or interactive for humans.
The loss function introduced above establishes a relativistic comparison mechanism, which encourages the generated structures to appear "more realistic than real ones" in a global average sense. 

Based on the structure-aware perceptual network obtained during the pretraining phase, we design a structure-aware perceptual loss (SAP loss). Unlike conventional perceptual networks, this design aims to preserve the structural information embedded in dense outputs such as those from modeling and RGT estimation. Accordingly, the perceptual network does not measure the similarity between the predictions and the labels directly. Instead, it computes the segmentation loss between these dense outputs and the structural labels, as follow,
\begin{equation}
	\mathcal{L}_\text{SAP}= \mathcal{L}_\text{seg}(\textbf{F}_\text{perc}(\hat{\textbf{Y}}_\text{dence}),\mathcal{S})
\end{equation}
where, \( \textbf{F}(\cdot) \) denotes the structure-aware perceptual network, $\hat{\textbf{Y}}_\text{dence}$ represents the dense prediction output generated by the generator, such as those produced in modeling and RGT estimation, and $\mathcal{S}$ denotes the structural label.

Due to the limitations of the training data available for the structure-aware perceptual network, it is less capable of capturing abstract semantic features in the manner of conventional LPIPS loss \cite{zhang2018unreasonable}. The SAP loss places greater emphasis on structural information present in seismic data. Therefore, to complement this limitation, the conventional LPIPS loss is also employed. As LPIPS is inherently a two-dimensional metric, a set of two-dimensional slices is randomly extracted along the three orthogonal directions of the output volume for the computation of LPIPS (Alex) loss .

The final loss of GEM is defined as follow,
\begin{equation}
	\mathcal{L}_\text{GEM} =  \alpha \cdot \mathcal{L}_\text{L1}+ 0.01 \cdot \mathcal{L}_\text{SAP} +  \mathcal{L}_\text{adv}+ \mathcal{L}_\text{LPIPS}
\end{equation}
where \( \alpha \) is scheduled using cosine annealing, decreasing smoothly from an initial value of 30 to a final value of $2.0$. This strategy gradually reduces the influence of voxel-level L1 supervision over training, allowing the model to increasingly emphasize adversarial, semantic-aligned, and perceptual consistency objectives.

The combination of the structure aware perceptual loss and relativistic adversarial training operationalizes physics informed objectives that promote geologically consistent and reliable predictions under limited supervision.

Training was performed on $8\times$  Nvidia H20 GPUs with a batch size of 8 for a total of 500,000 iterations. The first 300,000 iterations used an input size of \(320 \times 400 \times 128\) (timeline, inline, crossline), followed by 200,000 iterations with an increased size of \(512 \times 512 \times 128\) under PyTorch's checkpoint mode. During training, the inline and crossline dimensions were randomly transposed to improve spatial generalization.  The Adam optimizer is employed with the Two Time-Scale Update Rule (TTUR) \cite{heusel2017gans}. The learning rate is set to 0.001 for the generator and 0.004 for the discriminator, with \( \beta_1 = 0 \) and \( \beta_2 = 0.9 \).

\section{Conclusion}\label{sec13}
We introduce GEM, a prompt-based foundation model for subsurface imaging that unifies structural interpretation, stratigraphic analysis, geobody segmentation, and physical property modeling within a single generative framework. By reformulating all these tasks as prompt-guided conditional generation over structurally inferred latent spaces, GEM supports flexible human interaction and generalizes across surveys, data types, and geological settings without task-specific retraining. 

Unlike conventional models constrained to specific tasks and surveys, GEM leverages large-scale self-supervised pretraining and adversarial fine-tuning with heterogeneous prompts to achieve zero-shot generalization across a broad range of subsurface interpretation and modeling tasks. It demonstrates strong performance on real-world seismic and radar datasets, including the delineation of complex fault systems, Martian polar stratigraphy, full-volume seismic stratigraphy, diverse geobody segmentation, and multi-physical-property modeling guided by sparse well-log inputs. By integrating multimodal expert cues with structural priors, GEM serves as a new interface between geoscientific expertise and generative, prompt-driven reasoning in Earth science.

To our knowledge, GEM is the first generative foundation model in subsurface understanding that supports expert-in-the-loop interaction, prompt-guided reasoning, and zero-shot generalization across tasks, surveys, and data types. More broadly, this work establishes a new modeling paradigm for scientific AI in domains where data is structurally rich but semantically sparse, and where insight emerges from contextual, expert-informed interpretation. GEM exemplifies how generative models can be steered by structural priors, enabling interpretable, scalable, and collaborative modeling of the Earth’s interior. We envision this structure-aware, promptable approach extending to broader domains in scientific machine learning, including physical simulation, inverse problems, and multi-modal Earth system modeling. In this sense, GEM advances the aims of physics informed machine learning by enabling data efficient and trustworthy predictive modeling and by supporting data driven scientific discovery in geophysics.

\bibliography{sample}

\begin{thebibliography}{10}
\urlstyle{rm}
\expandafter\ifx\csname url\endcsname\relax
  \def\url#1{\texttt{#1}}\fi
\expandafter\ifx\csname urlprefix\endcsname\relax\def\urlprefix{URL }\fi
\expandafter\ifx\csname doiprefix\endcsname\relax\def\doiprefix{DOI: }\fi
\providecommand{\bibinfo}[2]{#2}
\providecommand{\eprint}[2][]{\url{#2}}

\bibitem{yu2024crustal}
\bibinfo{author}{Yu, P.} \emph{et~al.}
\newblock \bibinfo{journal}{\bibinfo{title}{Crustal permeability generated
  through microearthquakes is constrained by seismic moment}}.
\newblock {\emph{\JournalTitle{Nature communications}}}
  \textbf{\bibinfo{volume}{15}}, \bibinfo{pages}{2057} (\bibinfo{year}{2024}).

\bibitem{liu2025natural}
\bibinfo{author}{Liu, Q.} \emph{et~al.}
\newblock \bibinfo{journal}{\bibinfo{title}{Natural hydrogen in the
  volcanic-bearing sedimentary basin: Origin, conversion, and production
  rates}}.
\newblock {\emph{\JournalTitle{Science Advances}}}
  \textbf{\bibinfo{volume}{11}}, \bibinfo{pages}{eadr6771}
  (\bibinfo{year}{2025}).

\bibitem{zhang2024feasibility}
\bibinfo{author}{Zhang, Y.}, \bibinfo{author}{Jackson, C.} \&
  \bibinfo{author}{Krevor, S.}
\newblock \bibinfo{journal}{\bibinfo{title}{The feasibility of reaching
  gigatonne scale co2 storage by mid-century}}.
\newblock {\emph{\JournalTitle{Nature Communications}}}
  \textbf{\bibinfo{volume}{15}}, \bibinfo{pages}{6913} (\bibinfo{year}{2024}).

\bibitem{creasy2024co2}
\bibinfo{author}{Creasy, N.} \emph{et~al.}
\newblock \bibinfo{journal}{\bibinfo{title}{Co2 rock physics modeling for
  reliable monitoring of geologic carbon storage}}.
\newblock {\emph{\JournalTitle{Communications Earth \& Environment}}}
  \textbf{\bibinfo{volume}{5}}, \bibinfo{pages}{333} (\bibinfo{year}{2024}).

\bibitem{wang2024earthquake}
\bibinfo{author}{Wang, T.} \emph{et~al.}
\newblock \bibinfo{journal}{\bibinfo{title}{Earthquake forecasting from
  paleoseismic records}}.
\newblock {\emph{\JournalTitle{Nature communications}}}
  \textbf{\bibinfo{volume}{15}}, \bibinfo{pages}{1944} (\bibinfo{year}{2024}).

\bibitem{li2023earthquake}
\bibinfo{author}{Li, J.}, \bibinfo{author}{Zhu, W.}, \bibinfo{author}{Biondi,
  E.} \& \bibinfo{author}{Zhan, Z.}
\newblock \bibinfo{journal}{\bibinfo{title}{Earthquake focal mechanisms with
  distributed acoustic sensing}}.
\newblock {\emph{\JournalTitle{Nature Communications}}}
  \textbf{\bibinfo{volume}{14}}, \bibinfo{pages}{4181} (\bibinfo{year}{2023}).

\bibitem{li2022layered}
\bibinfo{author}{Li, C.} \emph{et~al.}
\newblock \bibinfo{journal}{\bibinfo{title}{Layered subsurface in utopia basin
  of mars revealed by zhurong rover radar}}.
\newblock {\emph{\JournalTitle{Nature}}} \textbf{\bibinfo{volume}{610}},
  \bibinfo{pages}{308--312} (\bibinfo{year}{2022}).

\bibitem{zhang2024buried}
\bibinfo{author}{Zhang, L.} \emph{et~al.}
\newblock \bibinfo{journal}{\bibinfo{title}{Buried palaeo-polygonal terrain
  detected underneath utopia planitia on mars by the zhurong radar}}.
\newblock {\emph{\JournalTitle{Nature Astronomy}}}
  \textbf{\bibinfo{volume}{8}}, \bibinfo{pages}{69--76} (\bibinfo{year}{2024}).

\bibitem{cui2023similar}
\bibinfo{author}{Cui, X.}, \bibinfo{author}{Li, Z.} \& \bibinfo{author}{Hu, Y.}
\newblock \bibinfo{journal}{\bibinfo{title}{Similar seismic moment release
  process for shallow and deep earthquakes}}.
\newblock {\emph{\JournalTitle{Nature Geoscience}}}
  \textbf{\bibinfo{volume}{16}}, \bibinfo{pages}{454--460}
  (\bibinfo{year}{2023}).

\bibitem{bergen2019machine}
\bibinfo{author}{Bergen, K.~J.}, \bibinfo{author}{Johnson, P.~A.},
  \bibinfo{author}{de~Hoop, M.~V.} \& \bibinfo{author}{Beroza, G.~C.}
\newblock \bibinfo{journal}{\bibinfo{title}{Machine learning for data-driven
  discovery in solid earth geoscience}}.
\newblock {\emph{\JournalTitle{Science}}} \textbf{\bibinfo{volume}{363}},
  \bibinfo{pages}{eaau0323} (\bibinfo{year}{2019}).

\bibitem{clubb2023himalayan}
\bibinfo{author}{Clubb, F.~J.} \emph{et~al.}
\newblock \bibinfo{journal}{\bibinfo{title}{Himalayan valley-floor widths
  controlled by tectonically driven exhumation}}.
\newblock {\emph{\JournalTitle{Nature Geoscience}}}
  \textbf{\bibinfo{volume}{16}}, \bibinfo{pages}{739--746}
  (\bibinfo{year}{2023}).

\bibitem{mousavi2022deep}
\bibinfo{author}{Mousavi, S.~M.} \& \bibinfo{author}{Beroza, G.~C.}
\newblock \bibinfo{journal}{\bibinfo{title}{Deep-learning seismology}}.
\newblock {\emph{\JournalTitle{Science}}} \textbf{\bibinfo{volume}{377}},
  \bibinfo{pages}{eabm4470} (\bibinfo{year}{2022}).

\bibitem{laurenti2024probing}
\bibinfo{author}{Laurenti, L.} \emph{et~al.}
\newblock \bibinfo{journal}{\bibinfo{title}{Probing the evolution of fault
  properties during the seismic cycle with deep learning}}.
\newblock {\emph{\JournalTitle{Nature communications}}}
  \textbf{\bibinfo{volume}{15}}, \bibinfo{pages}{10025} (\bibinfo{year}{2024}).

\bibitem{wu2019faultseg3d}
\bibinfo{author}{Wu, X.}, \bibinfo{author}{Liang, L.}, \bibinfo{author}{Shi,
  Y.} \& \bibinfo{author}{Fomel, S.}
\newblock \bibinfo{journal}{\bibinfo{title}{Faultseg3d: Using synthetic data
  sets to train an end-to-end convolutional neural network for 3d seismic fault
  segmentation}}.
\newblock {\emph{\JournalTitle{Geophysics}}} \textbf{\bibinfo{volume}{84}},
  \bibinfo{pages}{IM35--IM45} (\bibinfo{year}{2019}).

\bibitem{wu2023mtl}
\bibinfo{author}{Wu, W.} \emph{et~al.}
\newblock \bibinfo{journal}{\bibinfo{title}{Mtl-faultnet: Seismic data
  reconstruction assisted multitask deep learning 3-d fault interpretation}}.
\newblock {\emph{\JournalTitle{IEEE Transactions on Geoscience and Remote
  Sensing}}} \textbf{\bibinfo{volume}{61}}, \bibinfo{pages}{1--15}
  (\bibinfo{year}{2023}).

\bibitem{gao2021fault}
\bibinfo{author}{Gao, K.}, \bibinfo{author}{Huang, L.} \&
  \bibinfo{author}{Zheng, Y.}
\newblock \bibinfo{journal}{\bibinfo{title}{Fault detection on seismic
  structural images using a nested residual u-net}}.
\newblock {\emph{\JournalTitle{IEEE Transactions on Geoscience and Remote
  Sensing}}} \textbf{\bibinfo{volume}{60}}, \bibinfo{pages}{1--15}
  (\bibinfo{year}{2021}).

\bibitem{alaudah2019machine}
\bibinfo{author}{Alaudah, Y.}, \bibinfo{author}{Micha{\l}owicz, P.},
  \bibinfo{author}{Alfarraj, M.} \& \bibinfo{author}{AlRegib, G.}
\newblock \bibinfo{journal}{\bibinfo{title}{A machine-learning benchmark for
  facies classification}}.
\newblock {\emph{\JournalTitle{Interpretation}}} \textbf{\bibinfo{volume}{7}},
  \bibinfo{pages}{SE175--SE187} (\bibinfo{year}{2019}).

\bibitem{gao2024optimizing}
\bibinfo{author}{Gao, Z.}, \bibinfo{author}{Wang, K.}, \bibinfo{author}{Wang,
  Z.} \& \bibinfo{author}{Gao, J.}
\newblock \bibinfo{journal}{\bibinfo{title}{Optimizing seismic facies
  classification through differentiable network architecture search}}.
\newblock {\emph{\JournalTitle{IEEE Transactions on Geoscience and Remote
  Sensing}}} \textbf{\bibinfo{volume}{62}}, \bibinfo{pages}{1--12}
  (\bibinfo{year}{2024}).

\bibitem{yang2023multi}
\bibinfo{author}{Yang, J.}, \bibinfo{author}{Wu, X.}, \bibinfo{author}{Bi, Z.}
  \& \bibinfo{author}{Geng, Z.}
\newblock \bibinfo{journal}{\bibinfo{title}{A multi-task learning method for
  relative geologic , horizons, and faults with prior information and
  transformer}}.
\newblock {\emph{\JournalTitle{IEEE Transactions on Geoscience and Remote
  Sensing}}} \textbf{\bibinfo{volume}{61}}, \bibinfo{pages}{1--20}
  (\bibinfo{year}{2023}).

\bibitem{xu20243d}
\bibinfo{author}{Xu, Z.}, \bibinfo{author}{Li, K.}, \bibinfo{author}{Huang,
  Z.}, \bibinfo{author}{Yin, R.} \& \bibinfo{author}{Fan, Y.}
\newblock \bibinfo{journal}{\bibinfo{title}{3d salt body segmentation method
  based on multi-view co-regularization}}.
\newblock {\emph{\JournalTitle{IEEE Transactions on Geoscience and Remote
  Sensing}}}  (\bibinfo{year}{2024}).

\bibitem{muller2024deep}
\bibinfo{author}{Muller, A.~P.} \emph{et~al.}
\newblock \bibinfo{journal}{\bibinfo{title}{Deep-salt: Complete
  three-dimensional salt segmentation from inaccurate migrated subsurface
  offset gathers using deep learning}}.
\newblock {\emph{\JournalTitle{Geophysical Prospecting}}}
  \textbf{\bibinfo{volume}{72}}, \bibinfo{pages}{2186--2199}
  (\bibinfo{year}{2024}).

\bibitem{yang2024salt3dnet}
\bibinfo{author}{Yang, L.} \emph{et~al.}
\newblock \bibinfo{journal}{\bibinfo{title}{Salt3dnet: A self-supervised
  learning framework for 3d salt segmentation}}.
\newblock {\emph{\JournalTitle{IEEE Transactions on Geoscience and Remote
  Sensing}}}  (\bibinfo{year}{2024}).

\bibitem{gao2021channelseg3d}
\bibinfo{author}{Gao, H.}, \bibinfo{author}{Wu, X.} \& \bibinfo{author}{Liu,
  G.}
\newblock \bibinfo{journal}{\bibinfo{title}{Channelseg3d: Channel simulation
  and deep learning for channel interpretation in 3d seismic images}}.
\newblock {\emph{\JournalTitle{Geophysics}}} \textbf{\bibinfo{volume}{86}},
  \bibinfo{pages}{IM73--IM83} (\bibinfo{year}{2021}).

\bibitem{yu2021attention}
\bibinfo{author}{Yu, J.} \& \bibinfo{author}{Wu, B.}
\newblock \bibinfo{journal}{\bibinfo{title}{Attention and hybrid loss guided
  deep learning for consecutively missing seismic data reconstruction}}.
\newblock {\emph{\JournalTitle{IEEE Transactions on Geoscience and Remote
  Sensing}}} \textbf{\bibinfo{volume}{60}}, \bibinfo{pages}{1--8}
  (\bibinfo{year}{2021}).

\bibitem{saad2025f}
\bibinfo{author}{Saad, O.~M.} \& \bibinfo{author}{Alkhalifah, T.}
\newblock \bibinfo{journal}{\bibinfo{title}{F-siamesefwi: A novel deep-learning
  framework for multisource full-waveform inversion}}.
\newblock {\emph{\JournalTitle{Geophysics}}} \textbf{\bibinfo{volume}{90}},
  \bibinfo{pages}{R221--R230} (\bibinfo{year}{2025}).

\bibitem{wu2025does}
\bibinfo{author}{Wu, Y.} \& \bibinfo{author}{Ma, J.}
\newblock \bibinfo{journal}{\bibinfo{title}{How does neural network
  reparametrization improve geophysical inversion?}}
\newblock {\emph{\JournalTitle{Journal of Geophysical Research: Machine
  Learning and Computation}}} \textbf{\bibinfo{volume}{2}},
  \bibinfo{pages}{e2025JH000621} (\bibinfo{year}{2025}).

\bibitem{li2023self}
\bibinfo{author}{Li, Y.}, \bibinfo{author}{Alkhalifah, T.},
  \bibinfo{author}{Huang, J.} \& \bibinfo{author}{Li, Z.}
\newblock \bibinfo{journal}{\bibinfo{title}{Self-supervised pretraining vision
  transformer with masked autoencoders for building subsurface model}}.
\newblock {\emph{\JournalTitle{IEEE Transactions on Geoscience and Remote
  Sensing}}} \textbf{\bibinfo{volume}{61}}, \bibinfo{pages}{1--10}
  (\bibinfo{year}{2023}).

\bibitem{brown2020language}
\bibinfo{author}{Brown, T.} \emph{et~al.}
\newblock \bibinfo{journal}{\bibinfo{title}{Language models are few-shot
  learners}}.
\newblock {\emph{\JournalTitle{Advances in neural information processing
  systems}}} \textbf{\bibinfo{volume}{33}}, \bibinfo{pages}{1877--1901}
  (\bibinfo{year}{2020}).

\bibitem{touvron2023llama}
\bibinfo{author}{Touvron, H.} \emph{et~al.}
\newblock \bibinfo{journal}{\bibinfo{title}{Llama: Open and efficient
  foundation language models}}.
\newblock {\emph{\JournalTitle{arXiv preprint arXiv:2302.13971}}}
  (\bibinfo{year}{2023}).

\bibitem{bi2024deepseek}
\bibinfo{author}{Bi, X.} \emph{et~al.}
\newblock \bibinfo{journal}{\bibinfo{title}{Deepseek llm: Scaling open-source
  language models with longtermism}}.
\newblock {\emph{\JournalTitle{arXiv preprint arXiv:2401.02954}}}
  (\bibinfo{year}{2024}).

\bibitem{kirillov2023segment}
\bibinfo{author}{Kirillov, A.} \emph{et~al.}
\newblock \bibinfo{title}{Segment anything}.
\newblock In \emph{\bibinfo{booktitle}{Proceedings of the IEEE/CVF
  International Conference on Computer Vision}}, \bibinfo{pages}{4015--4026}
  (\bibinfo{year}{2023}).

\bibitem{feichtenhofer2022masked}
\bibinfo{author}{Feichtenhofer, C.}, \bibinfo{author}{Li, Y.},
  \bibinfo{author}{He, K.} \emph{et~al.}
\newblock \bibinfo{journal}{\bibinfo{title}{Masked autoencoders as
  spatiotemporal learners}}.
\newblock {\emph{\JournalTitle{Advances in neural information processing
  systems}}} \textbf{\bibinfo{volume}{35}}, \bibinfo{pages}{35946--35958}
  (\bibinfo{year}{2022}).

\bibitem{di2020accelerating}
\bibinfo{author}{Di, H.}, \bibinfo{author}{Truelove, L.}, \bibinfo{author}{Li,
  C.} \& \bibinfo{author}{Abubakar, A.}
\newblock \bibinfo{journal}{\bibinfo{title}{Accelerating seismic fault and
  stratigraphy interpretation with deep cnns: A case study of the taranaki
  basin, new zealand}}.
\newblock {\emph{\JournalTitle{The Leading Edge}}}
  \textbf{\bibinfo{volume}{39}}, \bibinfo{pages}{727--733}
  (\bibinfo{year}{2020}).

\bibitem{schuster2024review}
\bibinfo{author}{Schuster, G.~T.}, \bibinfo{author}{Chen, Y.} \&
  \bibinfo{author}{Feng, S.}
\newblock \bibinfo{journal}{\bibinfo{title}{Review of physics-informed
  machine-learning inversion of geophysical data}}.
\newblock {\emph{\JournalTitle{Geophysics}}} \textbf{\bibinfo{volume}{89}},
  \bibinfo{pages}{T337--T356} (\bibinfo{year}{2024}).

\bibitem{chen2021seismic}
\bibinfo{author}{Chen, Y.} \& \bibinfo{author}{Saygin, E.}
\newblock \bibinfo{journal}{\bibinfo{title}{Seismic inversion by hybrid machine
  learning}}.
\newblock {\emph{\JournalTitle{Journal of Geophysical Research: Solid Earth}}}
  \textbf{\bibinfo{volume}{126}}, \bibinfo{pages}{e2020JB021589}
  (\bibinfo{year}{2021}).

\bibitem{wu2023sensing}
\bibinfo{author}{Wu, X.} \emph{et~al.}
\newblock \bibinfo{journal}{\bibinfo{title}{Sensing prior constraints in deep
  neural networks for solving exploration geophysical problems}}.
\newblock {\emph{\JournalTitle{Proceedings of the National Academy of
  Sciences}}} \textbf{\bibinfo{volume}{120}}, \bibinfo{pages}{e2219573120}
  (\bibinfo{year}{2023}).

\bibitem{heir2024inversion}
\bibinfo{author}{Heir, A.}, \bibinfo{author}{Aghayev, S.},
  \bibinfo{author}{Tran, C.} \& \bibinfo{author}{Molder, A.}
\newblock \bibinfo{journal}{\bibinfo{title}{Inversion with stratigraphy-guided
  deep learning}}.
\newblock {\emph{\JournalTitle{Geophysics}}} \textbf{\bibinfo{volume}{89}},
  \bibinfo{pages}{R377--R386} (\bibinfo{year}{2024}).

\bibitem{tilke2022stratigraphic}
\bibinfo{author}{Tilke, P.} \emph{et~al.}
\newblock \bibinfo{title}{Stratigraphic forward modeler for artificial
  intelligence and machine learning workflows}.
\newblock In \emph{\bibinfo{booktitle}{Second EAGE Digitalization Conference
  and Exhibition}}, vol. \bibinfo{volume}{2022}, \bibinfo{pages}{1--5}
  (\bibinfo{organization}{European Association of Geoscientists \& Engineers},
  \bibinfo{year}{2022}).

\bibitem{kanfar2025intelligent}
\bibinfo{author}{Kanfar, R.}, \bibinfo{author}{Alali, A.},
  \bibinfo{author}{Tonellot, T.-L.}, \bibinfo{author}{Salim, H.} \&
  \bibinfo{author}{Ovcharenko, O.}
\newblock \bibinfo{journal}{\bibinfo{title}{Intelligent seismic workflows: The
  power of generative ai and language models}}.
\newblock {\emph{\JournalTitle{The Leading Edge}}}
  \textbf{\bibinfo{volume}{44}}, \bibinfo{pages}{142--151}
  (\bibinfo{year}{2025}).

\bibitem{gao2024foundation}
\bibinfo{author}{Gao, H.} \emph{et~al.}
\newblock \bibinfo{journal}{\bibinfo{title}{A foundation model enpowered by a
  multi-modal prompt engine for universal seismic geobody interpretation across
  surveys}}.
\newblock {\emph{\JournalTitle{arXiv preprint arXiv:2409.04962}}}
  (\bibinfo{year}{2024}).

\bibitem{liu2024foundation}
\bibinfo{author}{Liu, Q.} \& \bibinfo{author}{Ma, J.}
\newblock \bibinfo{journal}{\bibinfo{title}{Foundation models for geophysics:
  Review and perspective}}.
\newblock {\emph{\JournalTitle{arXiv preprint arXiv:2406.03163}}}
  (\bibinfo{year}{2024}).

\bibitem{sheng2025workflow}
\bibinfo{author}{Sheng, H.} \emph{et~al.}
\newblock \bibinfo{journal}{\bibinfo{title}{On the workflow, opportunities and
  challenges of developing foundation model in geophysics}}.
\newblock {\emph{\JournalTitle{arXiv preprint arXiv:2504.17384}}}
  (\bibinfo{year}{2025}).

\bibitem{sheng2024seismic}
\bibinfo{author}{Sheng, H.} \emph{et~al.}
\newblock \bibinfo{journal}{\bibinfo{title}{Seismic foundation model (sfm): a
  next generation deep learning model in geophysics}}.
\newblock {\emph{\JournalTitle{Geophysics}}} \textbf{\bibinfo{volume}{90}},
  \bibinfo{pages}{1--64} (\bibinfo{year}{2024}).

\bibitem{dou2021attention}
\bibinfo{author}{Dou, Y.}, \bibinfo{author}{Li, K.}, \bibinfo{author}{Zhu, J.},
  \bibinfo{author}{Li, X.} \& \bibinfo{author}{Xi, Y.}
\newblock \bibinfo{journal}{\bibinfo{title}{Attention-based 3-d seismic fault
  segmentation training by a few 2-d slice labels}}.
\newblock {\emph{\JournalTitle{IEEE Transactions on Geoscience and Remote
  Sensing}}} \textbf{\bibinfo{volume}{60}}, \bibinfo{pages}{1--15}
  (\bibinfo{year}{2021}).

\bibitem{bangs2023slow}
\bibinfo{author}{Bangs, N.~L.} \emph{et~al.}
\newblock \bibinfo{journal}{\bibinfo{title}{Slow slip along the hikurangi
  margin linked to fluid-rich sediments trailing subducting seamounts}}.
\newblock {\emph{\JournalTitle{Nature Geoscience}}}
  \textbf{\bibinfo{volume}{16}}, \bibinfo{pages}{505--512}
  (\bibinfo{year}{2023}).

\bibitem{foss20173d}
\bibinfo{author}{Foss, F.~J.}, \bibinfo{author}{Putzig, N.~E.},
  \bibinfo{author}{Campbell, B.~A.} \& \bibinfo{author}{Phillips, R.~J.}
\newblock \bibinfo{journal}{\bibinfo{title}{3d imaging of mars' polar ice caps
  using orbital radar data}}.
\newblock {\emph{\JournalTitle{The Leading Edge}}}
  \textbf{\bibinfo{volume}{36}}, \bibinfo{pages}{43--57}
  (\bibinfo{year}{2017}).

\bibitem{OpendTect}
\bibinfo{author}{{dGB}}.
\newblock \bibinfo{title}{Opendtect projects}.
\newblock \bibinfo{howpublished}{\url{https://terranubis.com/datalist/free}}.

\bibitem{ronneberger2015u}
\bibinfo{author}{Ronneberger, O.}, \bibinfo{author}{Fischer, P.} \&
  \bibinfo{author}{Brox, T.}
\newblock \bibinfo{title}{U-net: Convolutional networks for biomedical image
  segmentation}.
\newblock In \emph{\bibinfo{booktitle}{Medical image computing and
  computer-assisted intervention--MICCAI 2015: 18th international conference,
  Munich, Germany, October 5-9, 2015, proceedings, part III 18}},
  \bibinfo{pages}{234--241} (\bibinfo{organization}{Springer},
  \bibinfo{year}{2015}).

\bibitem{chen2018encoder}
\bibinfo{author}{Chen, L.-C.}, \bibinfo{author}{Zhu, Y.},
  \bibinfo{author}{Papandreou, G.}, \bibinfo{author}{Schroff, F.} \&
  \bibinfo{author}{Adam, H.}
\newblock \bibinfo{title}{Encoder-decoder with atrous separable convolution for
  semantic image segmentation}.
\newblock In \emph{\bibinfo{booktitle}{Proceedings of the European conference
  on computer vision (ECCV)}}, \bibinfo{pages}{801--818}
  (\bibinfo{year}{2018}).

\bibitem{9052469}
\bibinfo{author}{Wang, J.} \emph{et~al.}
\newblock \bibinfo{journal}{\bibinfo{title}{Deep high-resolution representation
  learning for visual recognition}}.
\newblock {\emph{\JournalTitle{IEEE Transactions on Pattern Analysis and
  Machine Intelligence}}} \textbf{\bibinfo{volume}{43}},
  \bibinfo{pages}{3349--3364}, \doiprefix\url{10.1109/TPAMI.2020.2983686}
  (\bibinfo{year}{2021}).

\bibitem{dou2022md}
\bibinfo{author}{Dou, Y.} \emph{et~al.}
\newblock \bibinfo{journal}{\bibinfo{title}{Md loss: Efficient training of 3-d
  seismic fault segmentation network under sparse labels by weakening anomaly
  annotation}}.
\newblock {\emph{\JournalTitle{IEEE Transactions on Geoscience and Remote
  Sensing}}} \textbf{\bibinfo{volume}{60}}, \bibinfo{pages}{1--14}
  (\bibinfo{year}{2022}).

\bibitem{li2024faultseg3d}
\bibinfo{author}{Li, Y.}, \bibinfo{author}{Wu, X.}, \bibinfo{author}{Zhu, Z.},
  \bibinfo{author}{Ding, J.} \& \bibinfo{author}{Wang, Q.}
\newblock \bibinfo{journal}{\bibinfo{title}{Faultseg3d plus: A comprehensive
  study on evaluating and improving cnn-based seismic fault segmentation}}.
\newblock {\emph{\JournalTitle{Geophysics}}} \textbf{\bibinfo{volume}{89}},
  \bibinfo{pages}{N77--N91} (\bibinfo{year}{2024}).

\bibitem{bi2021deep}
\bibinfo{author}{Bi, Z.}, \bibinfo{author}{Wu, X.}, \bibinfo{author}{Geng, Z.}
  \& \bibinfo{author}{Li, H.}
\newblock \bibinfo{journal}{\bibinfo{title}{Deep relative geologic time: a deep
  learning method for simultaneously interpreting 3-d seismic horizons and
  faults}}.
\newblock {\emph{\JournalTitle{Journal of Geophysical Research: Solid Earth}}}
  \textbf{\bibinfo{volume}{126}}, \bibinfo{pages}{e2021JB021882}
  (\bibinfo{year}{2021}).

\bibitem{wang2024cigchannel}
\bibinfo{author}{Wang, G.}, \bibinfo{author}{Wu, X.} \& \bibinfo{author}{Zhang,
  W.}
\newblock \bibinfo{journal}{\bibinfo{title}{cigchannel: A massive-scale 3d
  seismic dataset with labeled paleochannels for advancing deep learning in
  seismic interpretation}}.
\newblock {\emph{\JournalTitle{Earth System Science Data Discussions}}}
  \textbf{\bibinfo{volume}{2024}}, \bibinfo{pages}{1--27}
  (\bibinfo{year}{2024}).

\bibitem{wu2020deep}
\bibinfo{author}{Wu, X.}, \bibinfo{author}{Yan, S.}, \bibinfo{author}{Qi, J.}
  \& \bibinfo{author}{Zeng, H.}
\newblock \bibinfo{journal}{\bibinfo{title}{Deep learning for characterizing
  paleokarst collapse features in 3-d seismic images}}.
\newblock {\emph{\JournalTitle{Journal of Geophysical Research: Solid Earth}}}
  \textbf{\bibinfo{volume}{125}}, \bibinfo{pages}{e2020JB019685}
  (\bibinfo{year}{2020}).

\bibitem{dou2024contrasinver}
\bibinfo{author}{Dou, Y.}, \bibinfo{author}{Li, K.}, \bibinfo{author}{Lv, W.},
  \bibinfo{author}{Li, T.} \& \bibinfo{author}{Xiao, Y.}
\newblock \bibinfo{journal}{\bibinfo{title}{Contrasinver: Ultra-sparse label
  semi-supervised regression for multi-dimensional seismic inversion}}.
\newblock {\emph{\JournalTitle{IEEE Transactions on Geoscience and Remote
  Sensing}}}  (\bibinfo{year}{2024}).

\bibitem{wu2021deep}
\bibinfo{author}{Wu, X.}, \bibinfo{author}{Yan, S.}, \bibinfo{author}{Bi, Z.},
  \bibinfo{author}{Zhang, S.} \& \bibinfo{author}{Si, H.}
\newblock \bibinfo{journal}{\bibinfo{title}{Deep learning for multidimensional
  seismic impedance inversion}}.
\newblock {\emph{\JournalTitle{Geophysics}}} \textbf{\bibinfo{volume}{86}},
  \bibinfo{pages}{R735--R745} (\bibinfo{year}{2021}).

\bibitem{wu2022seismic}
\bibinfo{author}{Wu, B.}, \bibinfo{author}{Xie, Q.} \& \bibinfo{author}{Wu, B.}
\newblock \bibinfo{journal}{\bibinfo{title}{Seismic impedance inversion based
  on residual attention network}}.
\newblock {\emph{\JournalTitle{IEEE Transactions on Geoscience and Remote
  Sensing}}} \textbf{\bibinfo{volume}{60}}, \bibinfo{pages}{1--17}
  (\bibinfo{year}{2022}).

\bibitem{Seitzer2020FID}
\bibinfo{author}{Seitzer, M.}
\newblock \bibinfo{title}{{pytorch-fid: FID Score for PyTorch}}.
\newblock \bibinfo{howpublished}{\url{https://github.com/mseitzer/pytorch-fid}}
  (\bibinfo{year}{2020}).
\newblock \bibinfo{note}{Version 0.3.0}.

\bibitem{wang2004image}
\bibinfo{author}{Wang, Z.}, \bibinfo{author}{Bovik, A.~C.},
  \bibinfo{author}{Sheikh, H.~R.} \& \bibinfo{author}{Simoncelli, E.~P.}
\newblock \bibinfo{journal}{\bibinfo{title}{Image quality assessment: from
  error visibility to structural similarity}}.
\newblock {\emph{\JournalTitle{IEEE transactions on image processing}}}
  \textbf{\bibinfo{volume}{13}}, \bibinfo{pages}{600--612}
  (\bibinfo{year}{2004}).

\bibitem{goodfellow2020generative}
\bibinfo{author}{Goodfellow, I.} \emph{et~al.}
\newblock \bibinfo{journal}{\bibinfo{title}{Generative adversarial networks}}.
\newblock {\emph{\JournalTitle{Communications of the ACM}}}
  \textbf{\bibinfo{volume}{63}}, \bibinfo{pages}{139--144}
  (\bibinfo{year}{2020}).

\bibitem{tiandesigning}
\bibinfo{author}{Tian, K.} \emph{et~al.}
\newblock \bibinfo{title}{Designing bert for convolutional networks: Sparse and
  hierarchical masked modeling}.
\newblock In \emph{\bibinfo{booktitle}{The Eleventh International Conference on
  Learning Representations}}.

\bibitem{zhang2018unreasonable}
\bibinfo{author}{Zhang, R.}, \bibinfo{author}{Isola, P.},
  \bibinfo{author}{Efros, A.~A.}, \bibinfo{author}{Shechtman, E.} \&
  \bibinfo{author}{Wang, O.}
\newblock \bibinfo{title}{The unreasonable effectiveness of deep features as a
  perceptual metric}.
\newblock In \emph{\bibinfo{booktitle}{Proceedings of the IEEE conference on
  computer vision and pattern recognition}}, \bibinfo{pages}{586--595}
  (\bibinfo{year}{2018}).

\bibitem{jolicoeur2018relativistic}
\bibinfo{author}{Jolicoeur-Martineau, A.}
\newblock \bibinfo{journal}{\bibinfo{title}{The relativistic discriminator: a
  key element missing from standard gan}}.
\newblock {\emph{\JournalTitle{arXiv preprint arXiv:1807.00734}}}
  (\bibinfo{year}{2018}).

\bibitem{heusel2017gans}
\bibinfo{author}{Heusel, M.}, \bibinfo{author}{Ramsauer, H.},
  \bibinfo{author}{Unterthiner, T.}, \bibinfo{author}{Nessler, B.} \&
  \bibinfo{author}{Hochreiter, S.}
\newblock \bibinfo{journal}{\bibinfo{title}{Gans trained by a two time-scale
  update rule converge to a local nash equilibrium}}.
\newblock {\emph{\JournalTitle{Advances in neural information processing
  systems}}} \textbf{\bibinfo{volume}{30}} (\bibinfo{year}{2017}).

\bibitem{USGS}
\bibinfo{author}{{USGS}}.
\newblock \bibinfo{title}{The national archive of marine seismic surveys, u.s.
  geological survey}.
\newblock
  \bibinfo{howpublished}{\url{https://walrus.wr.usgs.gov/namss/search/}}.

\bibitem{SARIG}
\bibinfo{author}{{SARIG}}.
\newblock \bibinfo{title}{Resource and energy georeference databases, south
  australian resources information gateway}.
\newblock
  \bibinfo{howpublished}{\url{https://walrus.wr.usgs.gov/namss/search/}}.

\bibitem{NLOG}
\bibinfo{author}{{NLOG}}.
\newblock \bibinfo{title}{Dutch oil and gas portal, netherlands oil and gas
  exploration and production information}.
\newblock \bibinfo{howpublished}{\url{https://www.nlog.nl/en}}.

\bibitem{SEG}
\bibinfo{author}{{SEG}}.
\newblock \bibinfo{title}{Open data on the seg wiki, society of exploration
  geophysicists}.
\newblock \bibinfo{howpublished}{\url{https://wiki.seg.org/wiki/Open_data/}}.

\bibitem{ravi2024sam}
\bibinfo{author}{Ravi, N.} \emph{et~al.}
\newblock \bibinfo{journal}{\bibinfo{title}{Sam 2: Segment anything in images
  and videos}}.
\newblock {\emph{\JournalTitle{arXiv preprint arXiv:2408.00714}}}
  (\bibinfo{year}{2024}).

\bibitem{ma2025medsam2}
\bibinfo{author}{Ma, J.} \emph{et~al.}
\newblock \bibinfo{journal}{\bibinfo{title}{Medsam2: Segment anything in 3d
  medical images and videos}}.
\newblock {\emph{\JournalTitle{arXiv preprint arXiv:2504.03600}}}
  (\bibinfo{year}{2025}).

\bibitem{liu2022convnet}
\bibinfo{author}{Liu, Z.} \emph{et~al.}
\newblock \bibinfo{title}{A convnet for the 2020s}.
\newblock In \emph{\bibinfo{booktitle}{Proceedings of the IEEE/CVF conference
  on computer vision and pattern recognition}}, \bibinfo{pages}{11976--11986}
  (\bibinfo{year}{2022}).

\bibitem{woo2023convnext}
\bibinfo{author}{Woo, S.} \emph{et~al.}
\newblock \bibinfo{title}{Convnext v2: Co-designing and scaling convnets with
  masked autoencoders}.
\newblock In \emph{\bibinfo{booktitle}{Proceedings of the IEEE/CVF conference
  on computer vision and pattern recognition}}, \bibinfo{pages}{16133--16142}
  (\bibinfo{year}{2023}).

\bibitem{tan2019efficientnet}
\bibinfo{author}{Tan, M.} \& \bibinfo{author}{Le, Q.}
\newblock \bibinfo{title}{Efficientnet: Rethinking model scaling for
  convolutional neural networks}.
\newblock In \emph{\bibinfo{booktitle}{International conference on machine
  learning}}, \bibinfo{pages}{6105--6114} (\bibinfo{organization}{PMLR},
  \bibinfo{year}{2019}).

\bibitem{tan2021efficientnetv2}
\bibinfo{author}{Tan, M.} \& \bibinfo{author}{Le, Q.}
\newblock \bibinfo{title}{Efficientnetv2: Smaller models and faster training}.
\newblock In \emph{\bibinfo{booktitle}{International conference on machine
  learning}}, \bibinfo{pages}{10096--10106} (\bibinfo{organization}{PMLR},
  \bibinfo{year}{2021}).

\bibitem{woo2018cbam}
\bibinfo{author}{Woo, S.}, \bibinfo{author}{Park, J.}, \bibinfo{author}{Lee,
  J.-Y.} \& \bibinfo{author}{Kweon, I.~S.}
\newblock \bibinfo{title}{Cbam: Convolutional block attention module}.
\newblock In \emph{\bibinfo{booktitle}{Proceedings of the European conference
  on computer vision (ECCV)}}, \bibinfo{pages}{3--19} (\bibinfo{year}{2018}).

\end{thebibliography}
\newpage
\begin{appendices}
\appendixpage 
\section{Challenges in Subsurface Imaging Tasks}\label{apd1}

\begin{figure}[!h]
	\includegraphics[scale=0.44]{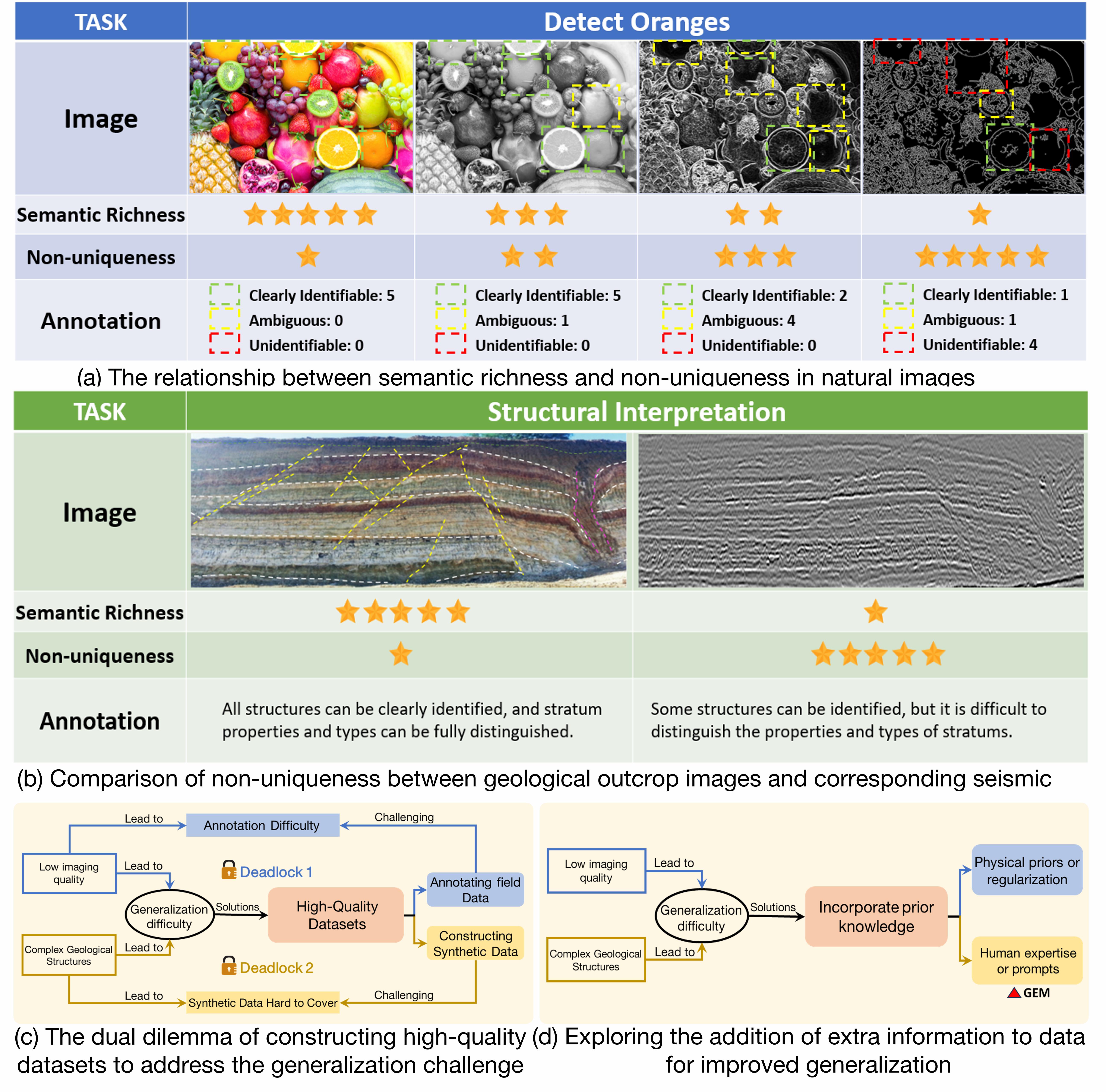}
	\centering\caption{Challenges and strategies for generalization in subsurface imaging tasks. (a) A visual analogy illustrates how removing color, texture, and gradient progressively reduces semantic information, increasing ambiguity for object recognition—mirroring the difficulty of subsurface interpretation from low-information seismic data. (b) Comparison between a geological outcrop and its corresponding seismic image. While structural features (e.g., faults and horizons) are clear in the outcrop, they become ambiguous or lost in seismic imagery, revealing the low information density of subsurface data. (c) Two current strategies for improving AI generalization: left, supervised learning from labeled field data faces annotation challenges due to low image quality; right, synthetic data generation offers better control but limited geological diversity. (d) A complementary strategy introduces physical priors or human expert guidance to augment the information content of seismic data—forming the basis for prompt-driven learning in GEM.}
	\label{fig06}
\end{figure}

Due to the low-quality nature of subsurface imaging information, providing explicit semantic representations for instances in seismic data is highly challenging. This is one of the key reasons why most related tasks struggle to generalize across different survey areas. We illustrate this issue with a step-by-step example.

In Figure \ref{fig06} (a), detecting oranges from the original image with rich semantic information is straightforward, aided by cues such as color, texture, gradients, and edges. As these elements are progressively removed—first color, then texture and gradient—the ambiguity of the image increases. Eventually, when only edge information remains, the task becomes highly uncertain and difficult. Identifying targets in such cases relies heavily on human prior knowledge. Applying AI methods to such inputs also significantly increases task complexity.

This challenge is even more pronounced in subsurface imaging. In Figure \ref{fig06} (b), structural interpretation of the geological outcrop on the left is relatively clear, with faults and horizons discernible and lithological variations interpretable from color differences. In contrast, the corresponding seismic data on the right obscures most structural features, rendering interpretation difficult for both humans and AI. Figure \ref{fig06} (c) (d) further illustrate two major strategies for addressing the generalization bottleneck in subsurface imaging tasks, along with their associated challenges.

\section{Training Data Preparation}\label{dataprep}
\begin{figure}[!h]
	\includegraphics[scale=0.25]{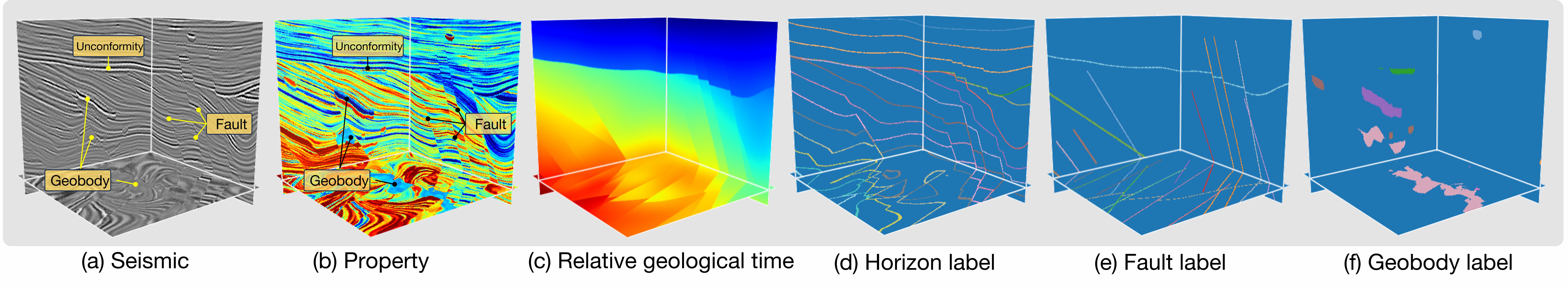}
	\centering\caption{Synthetic dataset with diverse labels constructed for fine-tuning GEM. (a) Seismic image containing key structural elements such as unconformities, faults, and geobodies. (b) Corresponding physical property volume (e.g., impedance), capturing internal stratigraphic and structural variability. (c) RGT volume, providing continuous stratigraphic ordering for full-horizon interpretation. (d–f). Semantic instance labels for horizons (d), faults (e), and geobodies (f). These property volumes, RGT maps, and semantic labels are used to supervise the fine-tuning of GEM, with diverse prompt types randomly sampled from these labels to guide the training across tasks. This dataset enables supervised fine-tuning with mixed prompt-label pairs across diverse subsurface imaging tasks, with spatially aligned, diverse labels embedded in a shared seismic volume. }
	\label{fig13}
\end{figure}

The training of GEM consists of two distinct stages. In the first stage, representation learning is conducted through self-supervised pretraining on massive unlabeled real-world 3D seismic imaging data from over 500 worldwide surveys. This stage enables the model to learn low-dimensional embeddings of structural features, thereby enhancing its ability to generalize across complex geological settings. In the subsequent stage, supervised fine-tuning is performed using accurately annotated synthetic datasets in order to adapt the pretrained model to specific interpretation or modeling tasks. As a result, the training data utilized by GEM comprises both real and synthetic components.

For field data, we mainly used the results collected by Sheng et al \cite{sheng2024seismic,USGS,SARIG,NLOG,SEG}.
These datasets span multiple geographic regions, including Central America, South Australia, Southeast Asia, and Northern Europe.
This globally sourced  data captures a wide range of representative subsurface geological features. These include various types of faults (such as normal, reverse, and strike-slip faults), folds exhibiting different degrees of structural complexity, geological bodies of diverse scales and spatial distributions, as well as multiple types and configurations of unconformities. The richness and diversity of these geological characteristics provide a valuable empirical basis for the analysis and interpretation of subsurface imaging data.

For synthetic data, we generated a dataset consisting of $1000$ $512\times512 \times512$ volumes with diverse types of labels, using forward modeling techniques. As shown in Figure \ref{fig13}, this synthetic samples include seismic data, acoustic impedance volumes, RGT (full horizon) annotations, as well as labeled faults and unconformities. In addition to the generated data, several publicly available synthetic datasets were incorporated during training \cite{wang2024cigchannel, wu2020deep}. These datasets include simulated representations of typical geological features such as fluvial channels and karstic caves, further enriching the diversity of geological scenarios encountered by the model.

\section{GEM's RGT Estimation Process}\label{rgtapdx}

\begin{figure}[!h]
	\includegraphics[scale=0.27]{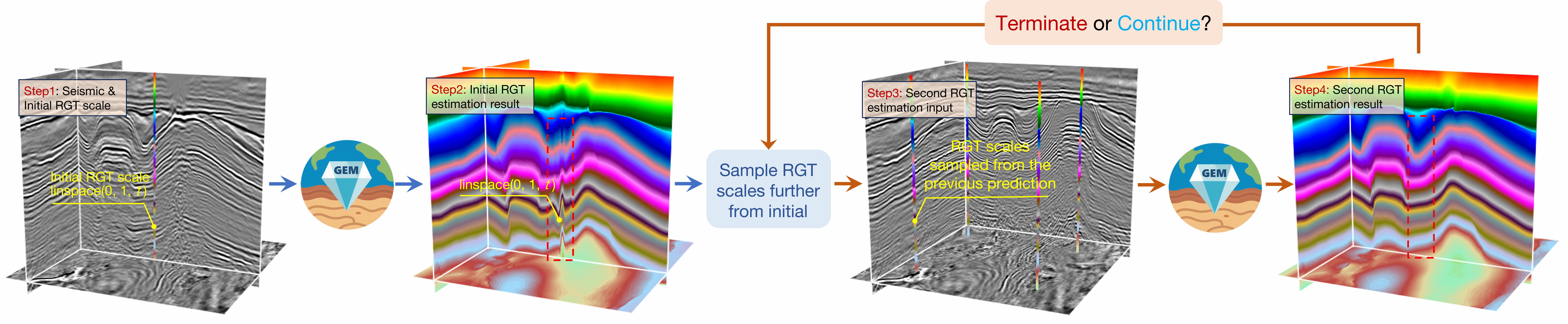}
	\centering\caption{Iterative RGT estimation using GEM. Step 1: An initial RGT scale uniformly distributed from 0 to 1 is aligned to the seismic volume center (pseudo-log style) and used as a prompt to GEM. Step 2: GEM produces an initial RGT prediction, which may contain misalignments due to scale uncertainty. Step 3: New RGT scales are sampled from regions where the initial prediction shows the largest deviation and are re-injected as updated prompts. Step 4: A second inference yields a refined RGT volume. If the prediction remains unreliable, the process is repeated until termination. This iterative mechanism enables fully automatic, prompt-based RGT estimation with self-correcting refinement.}
	\label{fig12}
\end{figure}

The estimation of RGT using GEM can be performed in a fully automatic manner, albeit requiring multiple rounds of inference. As illustrated in Figure \ref{fig12}, the process begins with an initial RGT scale of the same length as the timeline, uniformly distributed from 0 to 1. This initial scale is aligned to the seismic data volume center in the manner of a pseudo-log and jointly input into GEM. The first inference yields a preliminary RGT prediction, though the alignment of the RGT scale is typically inaccurate at this stage. To refine the result, four new RGT scales are selected from positions that deviate significantly from the initial scale and used as inputs for a second inference. The resulting prediction is then evaluated for reliability. In most cases, this second iteration already produces a reasonably accurate result. Depending on the outcome, the refinement process may be repeated or terminated.

\section{Additional Qualitative Experimental Results}
\subsection{Structural Interpretation}
\begin{figure}[!h]
	\includegraphics[scale=0.36]{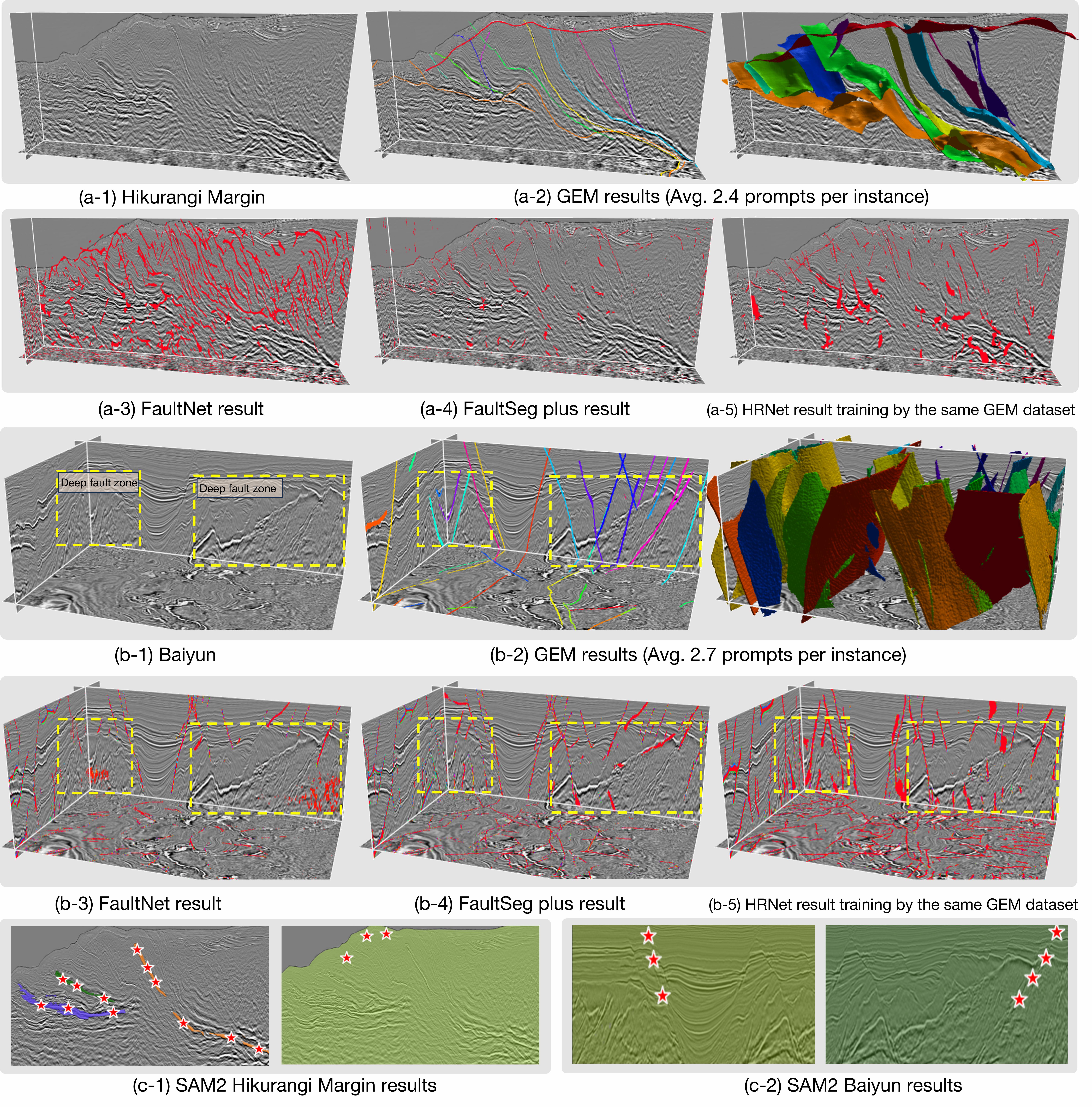}
	\centering\caption{
		Qualitative comparison of complex fault interpretation across different tectonic settings. (a, b) Fault interpretation results in the Hikurangi Margin (a) and Baiyun survey (b). GEM produces geologically coherent fault networks using an average of 2.4 and 2.7 prompts per 3D fault surface, respectively. (a-3) to (a-5) / (b-3) to (b-5). Results from FaultNet\cite{dou2022md}, FaultSeg+\cite{wu2019faultseg3d,li2024faultseg3d}, and HRNet (trained on the same dataset as GEM)—three widely used AI-based fault segmentation methods. While these models perform well in shallower normal fault zones (e.g., Baiyun), they struggle to generalize to more complex settings like thrusts and décollements in the Hikurangi Margin or deep-seated faults in Baiyun. (c) Interpretation results from the general-purpose segmentation model SAM2 on single 2D sections. Without geophysical fine-tuning, SAM2 fails to capture deep fault structures. Yellow bounding boxes highlight key fault zones of interest.}
	\label{fig07}
\end{figure}

In Figure \ref{fig02}, we present the performance of GEM alongside two open-source models—both of which have been integrated into multiple commercial software platforms \cite{dou2022md, wu2019faultseg3d, li2024faultseg3d}—and three conventional AI models, all trained on the same dataset, with quantitative results provided. Here, we focus on showcasing their qualitative performance.

The open-source models perform well in interpreting shallow, complex normal fault systems, as demonstrated in the Baiyun region shown in Figure \ref{fig07}. This is largely due to the fact that their training datasets are constructed based on similar data. However, as illustrated in Figure \ref{fig07}, these models struggle to generalize to other tectonic settings, such as the thrust and detachment faults in the Hikurangi margin and the deeper fault structures in Baiyun.

We also present fault interpretation results using the interactive segmentation foundation model SAM2 from a single 2D section. The results demonstrate that SAM2, when not fine-tuned for geophysical applications, cannot generalize effectively to subsurface imaging data.

\subsection{Geobody Segmentation}
\begin{figure}[!h]
	\includegraphics[scale=0.525]{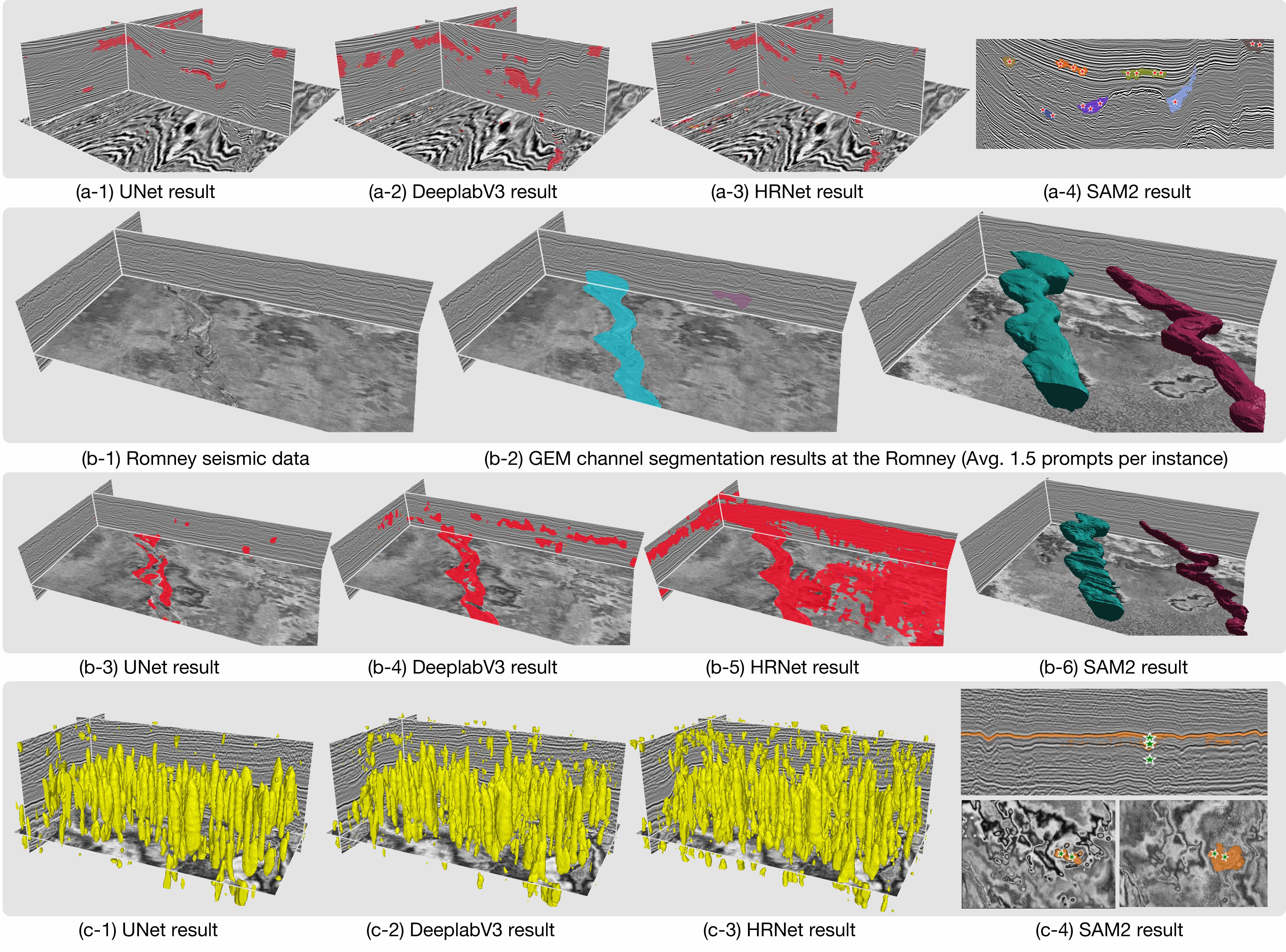}
	\centering\caption{Qualitative comparison of geobody segmentation for channels and karst caves. (a, b) Channel segmentation results on the Parihaka and Romney datasets. GEM achieves more complete and geologically coherent channel delineation using 3.5 (Parihaka) and 1.5 (Romney) prompts per 3D object on average. In contrast, UNet, DeepLabV3, and HRNet—trained on the same data as GEM—frequently miss key portions of the channel or introduce false positives. The SAM2 foundation model, adapted via 2D video-style extension, produces fragmented and discontinuous results due to its 2.5D nature, with visible stitching artifacts. (c) Karst segmentation comparisons on synthetic data. While traditional methods produce competitive results, GEM maintains structural continuity with fewer artifacts. SAM2 fails to capture the complex topology of karst features, even in interactive mode. These results highlight GEM’s strong spatial reasoning and zero-shot adaptability across distinct geological targets.}
	\label{fig08}
\end{figure}

Figure \ref{fig08} provides additional qualitative comparisons of geobody segmentation results across multiple methods, focusing primarily on channel and karst features—two synthetic geobody types present in our dataset. The first example illustrates channel detection in the Parihaka dataset, where all three baseline methods, despite being trained on the same data as GEM, exhibit significant omissions. In contrast, GEM achieves more complete segmentation. The interactive foundation model SAM2, when guided by a point-based prompt, is able to partially identify the channel on individual sections, but its results remain fragmented and incomplete.

In the karst segmentation task, traditional models perform effectively, achieving results comparable to or even surpassing those of GEM. In contrast, SAM2 fails to produce satisfactory segmentation results, even when operating in interactive mode.

In the Romney, SAM2 performs reasonably well. By adopting a video segmentation strategy, we used SAM2 to segment the complete channel body in the Romney seismic data, as shown in Figure \ref{fig08} (d-4). However, since SAM2 is inherently a 2D model and extends to 3D through temporal attention modules that sequentially link multiple 2D predictions, it essentially operates as a 2.5D model. This results in noticeable discontinuities and jagged artifacts along the stitching axis. In contrast, GEM produces smoother and more continuous results. A similar comparison can be observed in the VFS study \cite{gao2024foundation}, which employs many of the same real-world datasets as used in this work. These observations indicate that current 2D and 2.5D models based on SAM cannot effectively capture true 3D semantics and struggle to maintain spatial continuity in all directions.

\subsection{Property Modeling}

\begin{figure}[!h]
	\includegraphics[scale=0.222]{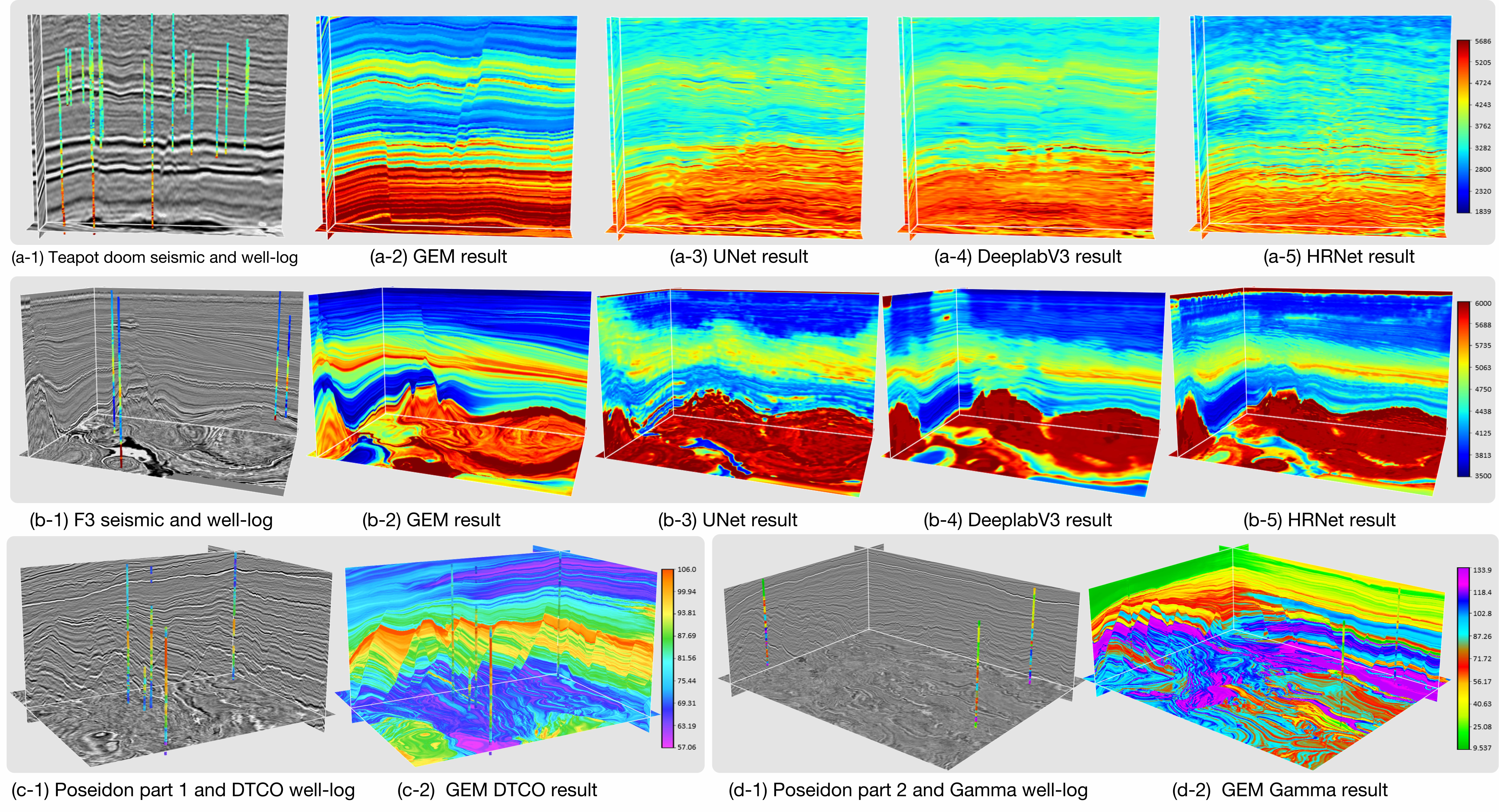}
	\centering\caption{
		Qualitative comparison of physical property modeling across diverse geological settings. (a, b) Acoustic impedance modeling on the Teapot Dome and F3 datasets using sparse well logs. Conventional AI models (UNet, DeepLabV3, HRNet) trained with task-specific supervision exhibit discontinuities and lateral instability, particularly on the F3 dataset with limited well coverage. In contrast, GEM produces laterally consistent impedance volumes using well logs as prompts, without task-specific retraining. (c) DTCO modeling in the structurally complex Poseidon dataset. Despite strong lateral variation and sparse well control, GEM captures fault displacements and unconformity terminations with high geometric fidelity. (d) Gamma ray modeling in the same region. These results highlight GEM’s robustness in physical property modeling, demonstrating strong structural awareness, lateral continuity, and adaptability across structurally complex surveys with limited well-log prompts.	}
	\label{fig09}
\end{figure}

Figure \ref{fig09} presents the impedance modeling results of conventional methods on the Teapot and F3 datasets. These traditional approaches uniformly adopt a multidimensional modeling strategy, wherein sparse 1D well-log labels are used to supervise 3D seismic volumes. Compared to the physical property modeling outcomes generated by GEM, the results from conventional impedance modeling exhibit insufficient lateral resolution and pronounced discontinuities along the reflectors. This issue is especially evident in the F3 dataset, where numerous anomalous value regions appear. The root cause lies in the fact that only four wells are available in the F3 block, with just three used for training, leading to severe overfitting and a lack of reliability in the network’s predictions.

Figure \ref{fig09} (c) and (d)  present a more challenging dataset. In the Poseidon survey, the fault system near the base (Near Top Plover) exhibits a highly structured set of normal faults with listric geometries, forming a complex pattern of fault blocks and grabens.

As shown in Figure \ref{fig09} (c-1), the DTCO well log exhibits strong lateral variations in its upper section. Combined with the complex lower fault system and sparse well coverage, these factors make modeling particularly difficult. Nevertheless, GEM is still able to produce reliable physical property modeling results. In (c-2), the predicted property volume reveals a fault system consistent with seismic imaging, clearly delineating the vertical termination of unconformities and their superposition on fault structures, thereby revealing the deformation history of stratums under tectonic processes. The predicted property responses exhibit strong geometric continuity along fault trends and accurately capture intense heterogeneity in fault intersection zones.

Figure \ref{fig09} (d-1) presents Gamma well logs, which are even more sparsely distributed. Yet in (d-2), GEM produces structurally coherent property volumes in geologically complex regions, faithfully reconstructing the spatial layout of the fault system and inter-stratal relationships. Even in zones of fault intersections and stratigraphic unconformities, the structural contours remain well defined.

This example demonstrates that GEM maintains strong structural awareness and expressive capacity in physical property modeling even under sparse well control. The results further validate the model’s robustness in parsing structural frameworks under prompt guidance, highlighting its generalization ability and practical potential in geologically complex and data-limited scenarios.

\section{Backbone Network Architecture}

\begin{figure}[!h]
	\includegraphics[scale=0.585]{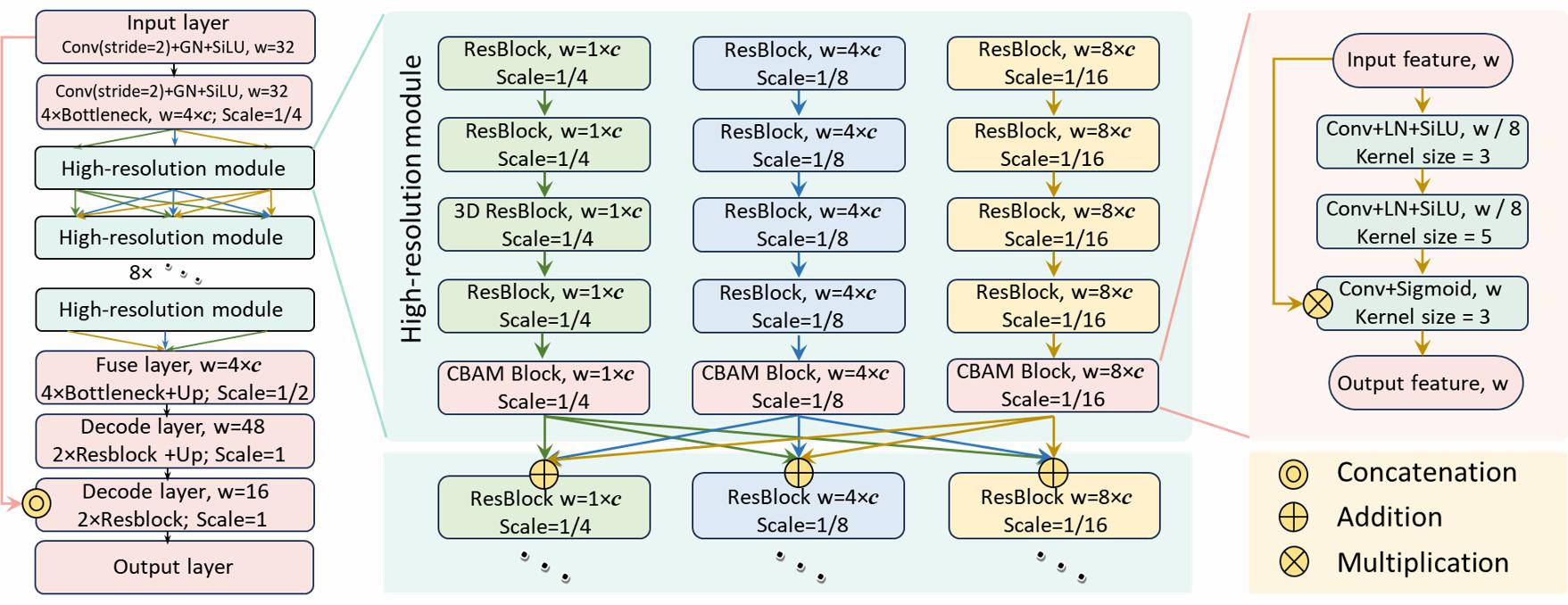}
	\centering\caption{
		Backbone network architecture of GEM. The GEM backbone is a pure convolutional architecture adapted from HRNet\cite{9052469}, designed to balance high-resolution feature preservation and semantic abstraction in 3D seismic volumes. The network maintains a persistent 1/4-scale high-resolution branch with moderate channel width to retain spatial detail, while two complementary low-resolution branches extract semantically rich representations. These branches receive repeated cross-scale feature injections from the high-resolution stream to mitigate information loss. A lightweight convolutional attention module—based on a modified CBAM—is integrated at each resolution to improve generalization by adaptively enhancing structural features. This architecture enables GEM to perform efficient and expressive 3D interpretation using only 0.67B parameters in its base configuration. }
	\label{fig10}
\end{figure}

For the backbone network, we adapt a classical pure convolutional architecture, HRNet \cite{9052469}, to accommodate 3D seismic data. Convolutional networks are well suited for constructing a genuinely 3D foundation model, as opposed to 2.5D approaches that merely apply 3D modules in parallel to two-dimensional feature extractors \cite{ravi2024sam,ma2025medsam2}. In contrast, employing transformer-based or self-attention architectures in 3D scenarios would substantially increase the computational cost of network construction, often exceeding the capabilities of current mainstream hardware.

Although a wide range of convolutional architectures is available, such as the widely adopted UNet \cite{ronneberger2015u}, ConvNeXt  \cite{liu2022convnet,woo2023convnext}, and the EfficientNet family \cite{tan2019efficientnet,tan2021efficientnetv2}, these models are typically designed following an encoder-decoder paradigm when applied to voxel-level tasks.
This presents a critical limitation when applied to 3D data. Specifically, they struggle to strike a balance between minimizing information loss and maintaining a manageable number of parameters.

In 3D data processing, each 2× downsampling operation results in an eightfold loss of spatial information. To preserve the total information throughput during feature propagation, the network must compensate by increasing the channel width by a factor of eight. However, increasing the width inevitably leads to a corresponding rise in the number of parameters. 
Modern autoencoder-based architectures commonly adopt downsampling ratios of 16× or even 32×, which implies that the network must maintain a channel width of at least \(8^4\) or \(8^5\) at the lowest resolution to ensure adequate information retention. Such a requirement results in a substantial increase in the model’s parameter count, thereby complicating training, exacerbating the risk of overfitting, and diminishing the overall efficiency of parameter utilization.

As illustrated in Figure \ref{fig10}, GEM's backbone preserves a 1/4-scale feature map throughout the forward pass, ensuring the retention of high-resolution information. Maintaining this branch requires only a moderate channel width of 64 ($8^2$). Building upon this foundation, we introduce modifications to the architecture by retaining only two low-resolution branches, which are responsible for capturing more abstract semantic representations. Although these branches inevitably suffer from information loss due to reduced spatial resolution, they repeatedly receive feature transmissions from the high-resolution branch during forward propagation. This cross-scale information exchange effectively compensates for the missing details.

In addition, we incorporated a convolutional attention module (CBAM) \cite{woo2018cbam} to enhance the network’s generalization capability through adaptive parameterization. As shown in Figure \ref{fig10}, unlike the original CBAM, we merged the spatial and channel attention mechanisms into a unified form, making the module more concise and computationally efficient.

These advantageous characteristics enable the GEM framework to achieve comprehensive structural interpretation and modeling in 3D subsurface imaging tasks, using only 0.67B parameters in its base configuration.

\section{Training and Inference Cost}

All training stages were carried out using eight NVIDIA H20 GPUs (96 GB each). The model comprises 0.67 billion parameters. During the self-supervised pretraining phase, the input size was $160\times160\times160$ with a batch size of 64. The model was trained for 500,000 steps, each taking approximately 1.8 seconds, totaling around 10 days.
For downstream tasks, the first 300,000 steps used an input size of $320\times400\times128$ with a batch size of 8, and each step took approximately 1.7 seconds. The remaining 200,000 steps were trained under PyTorch’s checkpoint mode with an input size of $512\times512\times128$ and a batch size of 8, requiring about 3.2 seconds per step, for a total of roughly 13 days.
During inference, processing a $512\times512\times128$ volume requires approximately $5.4 \times 10^4$ GFLOPs. At float16 precision, the memory usage is around 30 GB.

For interactive inference tasks such as structural interpretation and geobody segmentation, we recommend resizing the data to lower resolutions or smaller volumes on the GPU to improve interactivity. In contrast, tasks like physical property modeling and RGT estimation benefit from inference at the original resolution or volume size, which can be efficiently handled on CPU memory (CPU memory is more accessible and easier to scale, allowing for inference on larger volumes). This sacrifices inference speed but yields higher voxel-level accuracy and resolution.

\section{GUI for Interaction}
\begin{figure}[!h]
	\includegraphics[scale=0.45]{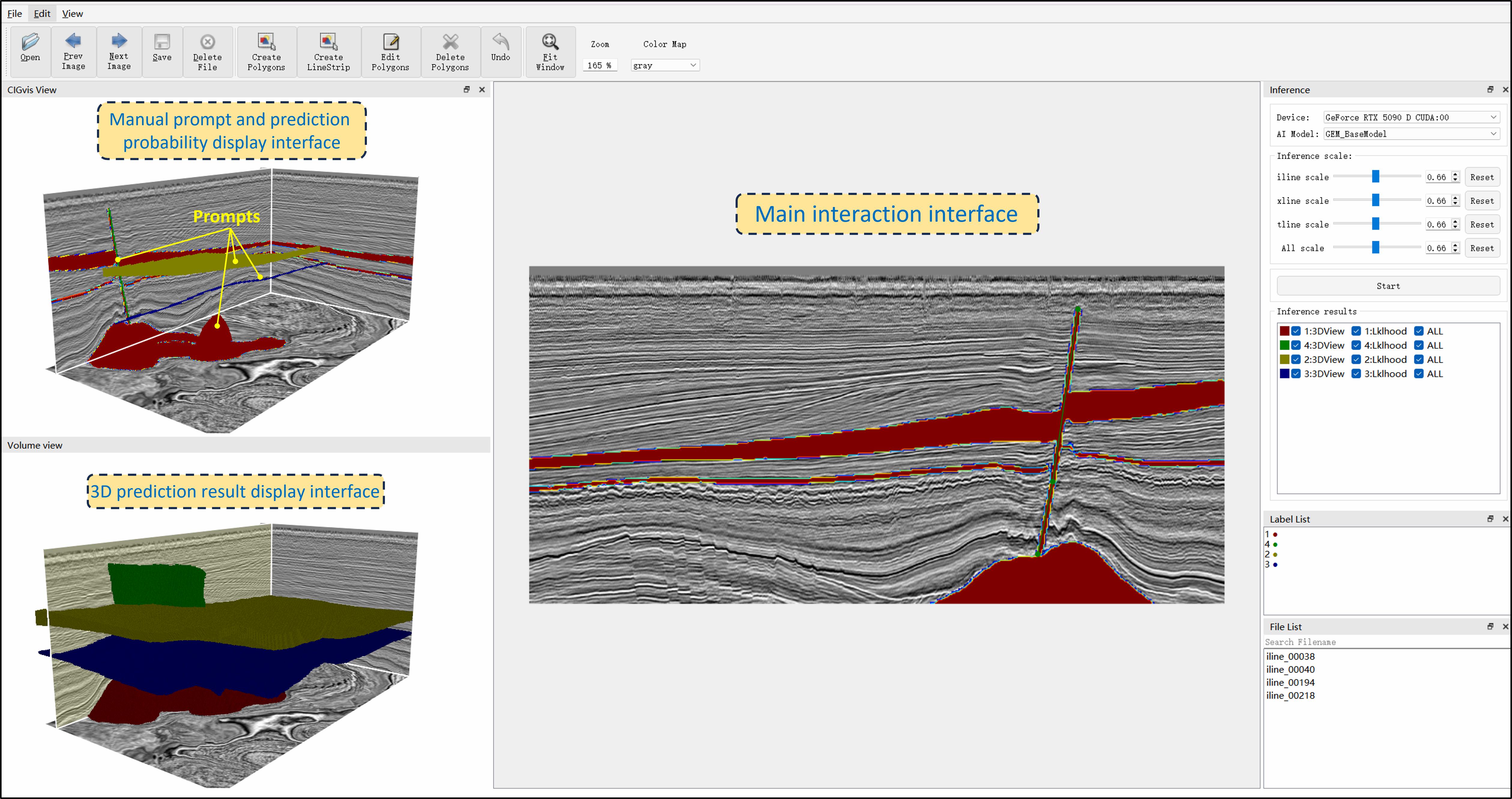}
	\centering\caption{This figure displays the graphical user interface (GUI) developed for interactive structural interpretation and geobody segmentation. The repository \url{https://douyimin.github.io/GEM} provides examples demonstrating how this interface is used for interactive interpretation.}
	\label{fig11}
\end{figure}

Figure \ref{fig11} shows the interactive GUI, more examples are provided on the accompanying project page.

\end{appendices}

\end{document}